\documentstyle[aps]{revtex}
\begin{document}
\newcommand{\beq}{\begin{equation}}
\newcommand{\eeq}{\end{equation}}
\newcommand{\beqn}{\begin{eqnarray}}
\newcommand{\eeqn}{\end{eqnarray}}
\newcommand{\bmath}{\begin{mathletters}}
\newcommand{\emath}{\end{mathletters}}
%\draft
\title{Superconductivity from Undressing. II. 
Single Particle Green's Function and Photoemission in Cuprates}
\author{J. E. Hirsch }
\address{Department of Physics, University of California, San Diego\\
La Jolla, CA 92093-0319}
 
\date{\today} 
\maketitle 
\begin{abstract} 
Experimental evidence indicates that the superconducting transition
in high $T_c$ cuprates is an 'undressing' transition. Microscopic
mechanisms giving rise to this physics were discussed in the
first paper of this series. Here we discuss the calculation of the
single particle Green's function and spectral function for Hamiltonians
describing undressing transitions in the normal and superconducting
states. A single parameter, $\Upsilon$, describes the strength of the 
undressing process and drives the  transition to superconductivity. 
In the normal state, the spectral function evolves from
predominantly incoherent to partly coherent as the hole concentration
increases. In the superconducting state,
the 'normal' Green's function acquires a contribution from the anomalous
Green's function when $\Upsilon$ is non-zero; the resulting contribution
to the spectral function is $positive$ for hole extraction and
$negative$ for hole injection. It is proposed that these results
explain the observation of sharp quasiparticle states in the
superconducting state of cuprates along the $(\pi,0)$ direction
and their absence along the $(\pi,\pi)$ direction.

\end{abstract}
\pacs{}

\section{Introduction}

Photoemission and optical experiments indicate that in high $T_c$ cuprates
a transition from an incoherent to a partially coherent state occurs
both as the hole doping increases in the normal state and as the system
goes superconducting\cite{ding,shen,puchkov,basov}. Basov and coworkers\cite{basov} 
have observed a lowering of c-axis kinetic energy as the transition to the
superconducting state occurs in several cuprates, especially in the
underdoped situation. It has been established however
that the $magnitude$ of the c-axis kinetic energy lowering 
detected is far too small to account for the superconducting condensation
energy at least in some cuprates\cite{marel}.  Ding, Campuzano, Norman and 
coworkers\cite{ding,campuzano,norman} as well as Feng and coworkers\cite{shen}
have reported observations of
sharp quasiparticle peaks in the superconducting state in
angle resolved photoemission emerging from a
highly incoherent normal state background along the $(\pi,0)$ direction,
close to the $(\pi/a,0)$ point.
 Ding and coworkers have 
interpreted the photoemission peak in terms of an enhanced
quasiparticle weight $Z$ in the superconducting state, and Feng and 
coworkers\cite{shen} have suggested that the peak in photoemission is a 
signature of the superfluid density.
Norman and coworkers\cite{norman} have analyzed the photoemission
observations in terms of a 'mode model' and emphasized the close 
connection between their observations and Basov's observation of
kinetic energy lowering. Furthermore, Basov and coworkers have emphasized that kinetic
energy lowering seems to occur only when there is a high degree
of incoherence in the normal state, and appears to vanish as the normal
state becomes more coherent (overdoped regime)\cite{basov}. They have furthermore
 proposed 
that the photoemission experiments suggest that kinetic energy lowering may occur
also $in-plane$ in the cuprates albeit only along the $(\pi,0)$ direction,
and for that reason may be difficult to observe directly.

The model of hole superconductivity\cite{holesc1,holesc2} $predicted$,
before the experimental observations, that
the superconducting condensation energy originates in in-plane kinetic
energy lowering\cite{area} and arises from a process of $undressing$
$of$ $hole$ $carriers$ as the pairing state develops\cite{color}. Thus it
describes both the kinetic energy lowering, arising from the low energy
effective Hamiltonian, as well as the high energy optical
spectral weight transfer,
that has also been observed experimentally\cite{fugol}.
In the first paper of this series\cite{undr} (hereafter referred to as I) we
formulated more generally the principles of superconductivity through
hole undressing, and pointed out that this physics would show up both in the one and
two-particle Green's functions, in qualitative agreement with the
observations reported above. Here we report calculation of the
single particle Green's function and spectral weight in the
superconducting state and discuss implications for the
understanding of photoemission  experiments.

\section{General principles}
In the class of models discussed in I, the wave function renormalization
of  quasiparticles is a function of the site occupation in
a local representation. The wavefunction renormalization arises from
coupling to a local boson degree of freedom. Three examples of specific
microscopic Hamiltonians describing this physics were discussed in I.
The 'coherent' part of the electron creation operator 
at site $i$ is defined by the following transformation:
\beq
d_{i\sigma}^\dagger=[T - (T-S)\tilde{n}_{di,-\sigma}]\tilde{d_{i\sigma}}^\dagger
\eeq
with $ 0 \leq S <T \leq 1$. The $\tilde{d} $ operators  in Eq. (1) are quasiparticle
operators\cite{nozieres}, and $\tilde{n}_d$ is the electron site
occupation. Eq. (1) expresses the fact that the electron
becomes less coherent as more electrons are added to the band. 
It should be kept in mind that the 'coherent part' of the
electron operator on the left side of Eq. (1) is not the full
electron creation operator, as it does not contain terms that give rise to
excited states of the boson degree of freedom\cite{undr}.

It will
be more useful to use hole operators rather than electron operators
throughout this paper; we stress however that the
discussion can be consistently carried out in electron as well as in
hole representation. In terms of hole operators, the coherent part
of the hole creation operator is
\beq
c_{i\sigma}^\dagger=[S + (T-S)\tilde{n}_{i,-\sigma}]\tilde{c_{i\sigma}}^\dagger
\equiv S[1 + \Upsilon\tilde{n}_{i,-\sigma}]\tilde{c_{i\sigma}}^\dagger
\eeq
Eq. (2) expresses the fact that the hole quasiparticle weight will 
increase with the local hole concentration, from $S$ in the regime of
low hole concentration to $T$ for high hole concentration. The high
degree of incoherence observed in high $T_c$ cuprates for low
hole doping implies $S<<1$, and the fact that coherence is 
achieved for relatively small values of hole doping implies
that the 'undressing parameter'
\beq
\Upsilon=\frac{T}{S}-1
\eeq
is very large. $\Upsilon$ is the parameter that drives the 
transition to the superconducting state. 
Note that a large $\Upsilon$ necessarily implies $S<<1$,
due to the constraint $T\leq 1$. For the normal state,
Eq. (2) implies for the hole  operator
\beq
c_{i\sigma}^\dagger= S[1 + \frac{n}{2}\Upsilon]\tilde{c_{i\sigma}}^\dagger
\eeq
with $n$ the hole concentration per site, and for the hole number
operator
\beq
n_{i\sigma}=S^2[1+\frac{n}{2}\Upsilon]^2\tilde{n}_{i\sigma}\equiv
Z(n)\tilde{n}_{i\sigma}
\eeq
with $Z(n)$ the hole quasiparticle weight. Eq. (5) implies
 that hole quasiparticles in the normal state become more coherent as the hole
concentration increases. In the limit $S\rightarrow 0$, 
quasiparticles become completely incoherent in the normal state for
low hole concentration and Fermi liquid theory breaks down. 
That limit is also described by the theory; in that limit,
the transition to the superconducting state is a superconductor-insulator
transition\cite{strong,bose}. Even though for that particular limiting
situation Fermi liquid theory does not describe the normal state,
we stress that  our
approach is $not$ a 'non-Fermi-liquid' approach, but instead is deeply
rooted in Fermi liquid theory.

Consider the bare kinetic energy in a tight binding model in terms
of hole operators
\beq
H_{kin}=-\sum_{i,j,\sigma}t_{ij}^0
(c_{i\sigma}^\dagger c_{j\sigma}+h.c.)
\eeq
Replacement of the bare hole operators by the quasiparticle
operators using eq. (2) yields
\bmath
\beq
H_{kin}=-\sum_{i,j,\sigma}t_{ij}^\sigma
(\tilde{c}_{i\sigma}^\dagger \tilde{c}_{j\sigma}+h.c.)
\eeq
\beq
t_{ij}^\sigma=t_{ij}^0 S^2[1+\Upsilon(\tilde{n}_{i,-\sigma}+\tilde{n}_{j,-\sigma})
+\Upsilon^2 \tilde{n}_{i,-\sigma}\tilde{n}_{j,-\sigma}]
\eeq
\emath
Eq. (7) expresses the fact that the hopping amplitude of a
hole quasiparticle will be increased, and as a consequence
its kinetic energy will be lowered, as the local hole
concentration increases; it is a direct consequence of the fact
that the quasiparticle coherence increases with local hole
concentration as described by Eq. (2). For low hole
concentration we can ignore the last term in Eq. (7b) and obtain
\bmath
\beq
H_{kin}=-\sum_{<i,j>,\sigma>}
[t_{ij}+\Delta t_{ij}(\tilde{n}_{i,-\sigma}+\tilde{n}_{j,-\sigma})]
(\tilde{c}_{i\sigma}^\dagger \tilde{c}_{j\sigma}+h.c.)
\eeq
\beq
t_{ij}=S^2t_{ij}^0
\eeq
\beq
\Delta t_{ij}=\Upsilon t_{ij}
\eeq
\emath
The kinetic energy of the form Eq. (8) is used in the model of hole
superconductivity, and leads to pairing and superconductivity
for low hole concentration in the presence of appreciable
on-site and nearest neighbor Coulomb repulsion\cite{holesc2}.
The condition for superconductivity to occur is
\beq
\Upsilon > \sqrt{(1+u)(1+w)}-1
\eeq
where $u$ and $w$ are dimensionless on-site and nearest neighbor
Coulomb repulsions\cite{holesc2}.
Hence within this class of models superconductivity is
intimately tied to increased quasiparticle coherence. Note that in a model
with anisotropy Eq. (8) still implies
\beq
\frac{\Delta t_{ij}}{t_{ij}}=\Upsilon
\eeq
$independent$ of direction. This assumption was used in our
studies with the model of hole superconductivity\cite{holesc2}, and can be
seen to be a necessary consequence of the fact that the
$\Delta t$ term in the Hamiltonian arises from  quasiparticle
undressing . A necessary consequence of Eq. (10) is
that the superconducting energy gap function has the form\cite{holesc2}
\beq
\Delta _k=\Delta (\epsilon_k)
\eeq
and hence is constant over the Fermi surface $(\epsilon_k=\epsilon_F)$,
even for an anisotropic band structure. Thus, Eq. (10)  can be 
understood as a direct consequence of the undressing physics. Finally,
Eq. (8) leads to superconductivity through kinetic energy
lowering\cite{area}. Hence, within the undressing scenario considered here,
kinetic energy lowering as the system goes superconducting is intimately
tied to  $s-wave$ $symmetry$ of the superconducting order parameter as described by
Eq. (11).

\section{Green's function: coherent part}
The single particle Green's function is given by a sum of coherent and incoherent parts
\beq
G_{rs}(\tau)=-<Tc_{r\uparrow}(\tau)c^\dagger_{s\uparrow}(0)>
\equiv G_{rs}^{coh}(\tau)+  G_{rs}^{incoh}(\tau)
\eeq
with $T$ the time ordering operator.
The coherent part of the Green's function is obtained by replacing the 
bare fermion operators in Eq. (12) by its coherent parts,
given by Eq. (2) in terms of the quasiparticle operators:
\beq
G_{rs}^{coh}(\tau)=
-S^2<T[1 + \Upsilon \tilde{n}_{r,\downarrow}(\tau)]\tilde{c}_{r\uparrow}(\tau)]
[1 + \Upsilon \tilde{n}_{s,\downarrow}(0)]\tilde{c}_{s\uparrow}^\dagger (0)]>
\eeq
The normal and anomalous Green's functions for the quasiparticle operators
\bmath
\beq
\tilde{G}_{rs}(\tau)=-<T\tilde{c}_{r\uparrow}(\tau)
\tilde{c}^\dagger_{s\uparrow}(0)>
\eeq
\beq
\tilde{F}_{rs}(\tau)=-<T\tilde{c}_{r\uparrow}(\tau)
\tilde{c}_{s\downarrow}(0)>
\eeq
\emath
are given by the usual form\cite{bcs}
\bmath
\beq
\tilde{G}(k,i\omega_n)=\frac{u_k^2}{i\omega_n-E_k}+
\frac{v_k^2}{i\omega_n+E_k}
\eeq
\beq\tilde{F}(k,i\omega_n)=-u_kv_k [\frac{1}{i\omega_n-E_k}-
\frac{1}{i\omega_n+E_k}]
\eeq
\emath
where the coherence factors $u_k$, $v_k$ and quasiparticle 
energies $E_k$ are given by the usual BCS expressions
\bmath
\beq
u_k^2=\frac{1}{2}(1+\frac{\epsilon_k-\mu}{E_k})
\eeq
\beq
v_k^2=\frac{1}{2}(1-\frac{\epsilon_k-\mu}{E_k})
\eeq
\beq
u_kv_k=\frac{\Delta_k}{2E_k}
\eeq
\beq
E_k=\sqrt{(\epsilon_k-\mu)^2+\Delta_k^2}
\eeq
\emath
and the 
gap function $\Delta_k$ is obtained from the BCS solution of
the model of hole superconductivity\cite{holesc2}, i.e. the
kinetic energy Eq. (8) supplemented with on-site and 
nearest neighbor Coulomb repulsion. The single particle energy
$\epsilon_k$ in these equations is given by $\epsilon_k=Z(n)\epsilon_k^0\sim
S^2(1+n\Upsilon)\epsilon_k^0$, with $\epsilon_k^0$ the bare kinetic
energy given by the Fourier transform of $(-t_{ij}^0)$.

 It can be seen that the extra density
operators in Eq. (13) will modify the normal Green's function introducing
anomalous terms, similar to the anomalous terms that occur when
calculating the expectation value of the kinetic energy Eq. (7) that
lead to the optical sum rule violation\cite{area}. We expand Eq. (13),
keeping only linear terms in the density
as appropriate to the low hole concentration regime, and use mean field
decoupling for the averages to obtain
\beq
G^{coh}(k,i\omega_n)=S^2[(1+n\Upsilon)\tilde{G}(k,i\omega_n)+
2f_0\Upsilon \tilde{F}(k,i\omega_n)]
\eeq
with $f_0=<\tilde{c}_{i\downarrow}\tilde{c}_{i\uparrow}>$
the on-site pair amplitude in the superconducting state. We have also performed
a space and time Fourier transform. It can be seen that the normal
Green's function has acquired a contribution from the anomalous Green's function
due to the density-dependent dressing.

However the quasiparticle spectral weights derived from Eq. (17) are not
positive definite and in fact can become negative in extreme parameter
regimes. To remedy this we need to include higher order terms obtained
from Eq. (13) by keeping terms with 6 fermion operators. Performing
a similar mean field decoupling for these we finally obtain for the
Green's function
\bmath
\beq
G^{coh}(k,i\omega_n)=\frac{ Z_h}{i\omega_n-E_k}+
\frac{ Z_e}{i\omega_n+E_k}
\eeq
with
\beq
Z_h=S^2  [[1+n\Upsilon]u_k-f_0\Upsilon v_k]^2
\eeq
\beq
Z_e=S^2  [[1+n\Upsilon]v_k+f_0\Upsilon u_k]^2
\eeq
and
\beq
f_0=<\tilde{c}_{i\downarrow}\tilde{c}_{i\uparrow}>=
\frac{1}{N}\sum_k \frac{\Delta_k}{2E_k}(1-2f(E_k)) .
\eeq
\emath
The quasiparticle weights $Z_h$ and $Z_e$ are clearly positive definite.
Their sum is not conserved as function of density or temperature
because of contributions from the
incoherent part of the Green's function not contained in Eq. (13).

In the absence of undressing ($\Upsilon=0$) the coherent Green's
function Eq. (18) reduces to the usual BCS form except for the
overall factor $S^2$.
In the presence of undressing ($\Upsilon>0$)
Eq. (18) shows that the coherent part of the Green's function
and spectral function will increase with hole density $n$, both for positive
and negative energies. Furthermore, as the system goes superconducting
the on-site pair amplitude $f_0$ develops a positive expectation
value. From Eq. (18) this implies that the coherent spectral
weight will $decrease$ for $positive$ energies (hole injection) and
$increase$ for $negative$ energies (hole extraction). The effect in
the superconducting state will be largest for parameters where the
on-site pair amplitude is large, which corresponds to short coherence
length achieved in the strong coupling underdoped regime\cite{strong}. The 
magnitude
of these effects both in the normal and superconducting state depend
on the magnitude of the undressing parameter $\Upsilon$. We discuss the
implications of these results in subsequent sections.

\section{Results for quasiparticle weights}

To illustrate the behavior emerging from the results of the previous
section we consider now a specific example. The quasiparticle Hamiltonian
is given by the kinetic energy Eq. (8) supplemented by on-site and
nearest neighbor Coulomb repulsion
\beq
H_{Coul}=U\sum_i \tilde{n}_{i\uparrow} \tilde{n}_{i\downarrow}
+V\sum_{<ij>}\tilde{n}_i\tilde{n}_j
\eeq
The BCS solution of this Hamiltonian\cite{holesc2} yields the quasiparticle
energies
\bmath
\beq
E_k=\sqrt{(\epsilon_k-\mu)^2+\Delta_k^2}= 
\sqrt{a^2(\epsilon_k-\mu-\nu)^2+\Delta_0^2}
\eeq
\beq
\Delta_k\equiv \Delta(\epsilon_k)=\Delta_m(-\frac{\epsilon_k}{D/2}+c)
\eeq
\beq
\Delta_0=\frac{1}{a}\Delta(\mu)
\eeq
\beq
\nu=\frac{1}{a}\frac{\Delta_m}{D/2}\Delta_0
\eeq
\beq
a=\frac{1}{\sqrt{1+(\frac{\Delta_m}{D/2})^2}}
\eeq
\emath
with $\Delta_m$ and $c$ parameters that depend on temperature and
doping. The bandwidth $D$ in these equations is given by
\beq
D=D_h(1+n\Upsilon)
\eeq
with $D_h$ the bandwidth in the limit of zero hole concentration.
The quasiparticle gap, i.e. the minimum quasiparticle excitation energy,
is given by
\beq
E_g=\Delta_0
\eeq
and occurs at momentum defined by
\beq
\epsilon_k^{[0]}=\mu+\nu
\eeq
However, if $\epsilon_k^{[0]}$ is below the bottom of the band, which occurs when
the chemical potential is sufficiently below the bottom of the band at low
hole concentration, Eq. (22) is not valid, and instead
\beq
E_g=\sqrt{(-\frac{D}{2}-\mu)^2+\Delta(-\frac{D}{2})}
\eeq

We consider a two-dimensional square lattice with only nearest neighbor hopping
and $t_{ij}=t_h$ in Eq. (8). The quasiparticle bandwidth as the hole
concentration goes to zero is $D_h=2zt_h$, with $z=4$ the number of
nearest neighbors to a site. We choose parameters
\beqn
D_h=0.2 eV  \nonumber \\
 U=5eV \nonumber \\
V=0.65 eV  \nonumber \\
\Upsilon = 19.2
\eeqn
which imply $\Delta t=\Upsilon D_h/2z=0.48 eV$. For the present purposes we
need not specify the magnitude of the parameter $S^2$, which determines
the relative weight of coherent and incoherent contributions to the spectral
function.

These parameters yield a maximum $T_c$ versus hole concentration of
$T_c^{max}=94K$, as shown in Fig. 1(a), for optimal doping $n\sim 0.045$. The 
minimum quasiparticle excitation energy at low temperatures
is shown in Fig. 1(b). At low hole concentration it does not go to zero as $T_c$
does because the chemical potential falls below the bottom of the
band and $E_g$ is determined by Eq. (24) rather than by Eq. (22). The behavior
of the chemical potential and the bottom of the band versus hole concentration
is shown in Fig. 2. The chemical potential crosses the
bottom of the band at $n\sim 0.038$, and $\epsilon_k^0$ (Eq. (23)) crosses the
bottom of the band at $n\sim 0.034$.

The on-site pair amplitude $f_0$ that enters in the expressions for the
quasiparticle weights is shown in Fig. 3. As function of doping it
follows approximately the behavior of the critical temperature and of the
gap parameter $\Delta_0$ (not shown). At low hole doping it goes
to zero because the carrier concentration goes to zero, at high hole
doping it goes to zero because the coherence length is diverging\cite{strong}. 
As function of temperature, $f_0$ behaves approximately like the gap,
going to zero at $T_c$ as $(T_c-T)^{1/2}$.

Next we consider the behavior of the quasiparticle weights $Z_e$ and
$Z_h$ as function of temperature. Figure 4 shows the results at the
(normal state) Fermi energy, $\epsilon_k=\mu$, for the optimally
doped case ($n=0.045$). The values are normalized so that $Z_e$ and
$Z_h$ would be $0.5$ for $\Upsilon=0$. The dashed line shows the value
the weights would have for $f_0=0$: it is temperature independent
and larger than $0.5$ because of the undressing due to the
average carrier concentration $n$. The effect of onset of superconductivity
is to increase $Z_e$ as the temperature is lowered and to decrease
$Z_h$. This indicates that there is extra amplitude for electron creation, 
and less amplitude for hole creation. It may thus be interpreted as a shift of the
chemical potential as superconductivity sets in, giving increased
hole occupation as the temperature is lowered, or equivalently 
a shrinking of the electron Fermi sea. This is a surprising result 
of this calculation,
and its implications will be discussed in  subsequent sections. Note
that the weight $Z_e$ increases by almost a factor of $2$ between
$T=T_c$ and $T=0$. The magnitude of the increase of course depends on the
magnitude of the undressing parameter $\Upsilon$, and would be larger
or smaller for larger or smaller values of $\Upsilon$ respectively.
By adjusting the values of on-site and nearest neighbor Coulomb
repulsion in the model it would be possible to obtain the same
maximum $T_c$ with different values of $\Upsilon$, as discussed
in previous work\cite{holesc2}. Nevertheless we believe that the parameters
chosen for this example may be representative of the situation in
high $T_c$ materials.

Note also that the total weight of the spectral function $Z_{tot}=Z_e+Z_h$
increases as the temperature is lowered below $T_c$. This
indicates that overall there is more coherence in the superconducting
than in the normal state, in accordance with qualitative expectations,
and this extra spectral weight is transfered from the high energy 
incoherent part of the spectral function as will be discussed in the
following section. However part of the enhancement of
$Z_e$ at low temperatures relative to its value at $T_c$ can be
attributed to spectral weight being transfered from negative to
positive energies (i.e. corresponding depletion of $Z_h$) in addition to
spectral weight transfer from the incoherent part of the spectral function.

Similarly Fig. 5 shows the results for an overdoped case, $n=0.1$, with $T_c=68K$.
The behavior is qualitatively similar as the optimally doped case,
however the effect of the onset of superconductivity on the spectral weights
is considerably smaller because the system is already more coherent in the
normal state. This is indicated by the larger values of all the spectral
weights relative to the values of the case shown in Fig. 4, due to the
enhanced coherence arising from the increased hole concentration.
For a much higher hole concentration, as $T_c$ approaches zero,
the 'gap' between $Z_e$ and $Z_h$ in the superconducting state closes,
as shown in Fig. 6. It remains always nonzero however as long as
$T_c$ is nonzero, and there is always some spectral weight transfer from
the incoherent region as long as $T_c$ is nonzero.

Next we consider the spectral weights for other values of momentum.
Figure 7 shows results for $\epsilon_k-\mu>0$. Recall that we are
using hole representation, so $\epsilon_k-\mu>0$ means inside the filled Fermi
sea for electrons. In the normal state $Z_e=0$: since the electron state
is full, no new electron can be created in it. Just as in the conventional 
BCS case, as superconductivity
sets in the state becomes partially occupied and  $Z_e\neq 0$ , and
$Z_h$ correspondingly decreases. However, unlike conventional BCS, 
$Z_e$ and $Z_h$ cross in our case and at low temperatures
the weight for creating an electron is larger than that for creating
a hole, even though we are inside the filled normal state Fermi sea. Clearly
this implies that the chemical potential in the superconducting state
has changed.
Figure 7(b) shows that for $\epsilon_k-\mu=16.1meV$ the weights for
electrons and holes coincide at low temperatures; this momentum then
corresponds to the new Fermi momentum, $k_F'$, in the superconducting
state. For even larger $\epsilon_k$, $Z_e$ becomes smaller than $Z_h$ as
in the conventional case, as shown in Fig. 7(c).

The behavior for negative energy (outside of the electron Fermi sea)
is shown in Fig. 8. Here, $\epsilon_k-\mu$ was chosen to be at the
bottom of the hole band in the optimally doped case. The weight for
electron creation is much larger than in the conventional case.
Note also that $Z_e$ first decreases and then increases as the
temperature is lowered, and as $T\rightarrow 0$ it becomes even
larger than its value in the normal state. Such a situation, 
which is never seen in the conventional case, is possible here due to the 
non-conservation of $Z_{tot}$ because of the transfer of spectral weight from
the incoherent part to the coherent part of the spectral function as
the temperature is lowered.

Figure 9 shows the spectral weights for an overdoped case, $n=0.1$,
for values of the momentum above the electron Fermi surface, at the Fermi
surface and below the Fermi surface. The behavior is qualitatively
similar to that for the optimally doped case, although the differences
between the conventional and our case are less pronounced here because
this is a weaker coupling regime. In figure 10 we show the spectral
weight for an underdoped case, $n=0.02$. Here the chemical potential is 
below the bottom of the band, so the situation comparable to Fig. 4 cannot be
attained. Fig. 10 shows the behavior of the spectral weight for $\epsilon_k$
at its lowest possible value, the bottom of the band, which is 
qualitatively similar to other cases where $\epsilon_k$ is above
$\mu$ such as Fig. 7(a).

Next we consider the behavior of the quasiparticle weights at the 
chemical potential versus doping in Fig. 11. The upper dot-dashed line is
the total spectral weight in the superconducting state, and the dotted
line below it is the total spectral weight in the normal state. The 
difference between the two is the spectral weight transfered from
high energy incoherent processes as the system goes superconducting; this
difference approaches zero in the 
overdoped regime. The full lines denote the quasiparticle weights in
our case, the dashed lines the usual BCS results ($u_k^2=Z_h, v_k^2=Z_e$),
which increase approximately linearly with $n$ due to the normal state 
increased coherence
with doping and are equal for $\epsilon_k=\mu$. For low dopings however
the chemical potential falls below the
bottom of the band and hence we take $\epsilon_k$  at the bottom of the
band rather than at $\mu$, this is why the two dashed lines diverge at
low dopings. In our case, the weight for electron creation 
(solid curve labeled $Z_e$) is seen to increase
rapidly with doping and then taper off for high doping; this latter effect
is due to the reduction of the on-site pair amplitude 
$f_0$ for high doping as the coherence length
becomes large\cite{holesc2}. The quasiparticle weights $Z_e$ and $Z_h$ approach
each other and the BCS value for high doping, as expected. Note also that there
is a narrow doping regime where the electron weight $Z_e$ is even larger
than the total weight in the normal state (dotted line). This situation can
never  occur in the conventional BCS case.

We believe the behavior exhibited by $Z_e$ in Fig. 11 is relevant to the
understanding of the angle resolved photoemission results discussed by
Ding et al\cite{ding}. In their work, the quasiparticle weight in the
superconducting state extracted from photoemission spectra shows
similar qualitative behavior to the behavior exhibited by $Z_e$ in 
Fig. 11. We will discuss the relation between $Z_e$ and the experimental
quantity in a subsequent section. Ding et al also plot $Z\Delta$, the product of 
their extracted quasiparticle weight and the gap inferred from the
photoemission spectra, and point out that its behavior rougly 
follows the bell-shaped curve of $T_c$. Our calculation shows similar
behavior, as shown in Fig. 12. Note that the quasiparticle gap itself
remains finite in our calculation as the hole concentration goes
to zero\cite{strong} (Fig. 1 (b)), and also experimentally\cite{gray}.

\section{Green's function: incoherent part}
To calculate the incoherent part of the Green's function we now
consider a specific model, the generalized Holstein model discussed
in I. Our calculation follows closely the calculation of
Alexandrov and Ranninger\cite{alex} for the conventional Holstein model, 
and we refer the reader to their seminal work for details which
are common to both situations. The site
Hamiltonian for our case is given by\cite{undr}
\beq
H=\hbar \omega_0 a^\dagger a +g \hbar \omega_0 (a^\dagger +a)
(n_\uparrow +n_\downarrow -\gamma n_\uparrow n_\downarrow)+
U n_\uparrow n_\downarrow
\eeq
The case of Alexandrov and Ranninger 
corresponds to $\gamma=0$. Using a generalized Lang-Firsov
transformation\cite{undr}
 the quasiparticle (polaron) operators $\tilde{c}_{i\sigma}$ are related
to the bare fermion (hole) operators by
\bmath
\beq
c_{i\sigma}=\tilde{c}_{i\sigma}X_{i\sigma}
\eeq
\beq
X_{i\sigma}=e^{-g(a_i^\dagger -a_i)(1-\gamma \tilde{n}_{i-\sigma})}
\eeq
\emath
In contrast to Eq. (2), the operator $c_{i\sigma}$ here is the full
hole destruction operator, including coherent and incoherent parts.
The coherent part results from
the expectation value of the $X$-operators in the zero boson subspace,
\beq
<X_{i\sigma}>=e^{-\frac{g^2}{2}(1-\gamma \tilde{n}_{i-\sigma})}
\eeq
and in particular
\bmath
\beq
X_{i\sigma}(\tilde{n}_{i-\sigma}=0)=e^{-\frac{g^2}{2}}=S
\eeq
\beq
X_{i\sigma}(\tilde{n}_{i-\sigma}=1)=e^{-\frac{g^2}{2}(1-\gamma)^2}=T
\eeq
\emath
in accordance with Eq. (2).

We wish to calculate the Green's function
\beqn
G(m,\tau)&=&-<Tc_{0\uparrow}(\tau)c^\dagger_{m\uparrow}(0)>=\nonumber \\
&=&-<Te^{g(a_0^\dagger(\tau)-a_0(\tau))(1-\gamma\tilde{n}_{0\downarrow}(\tau))}
\tilde{c}_{0\uparrow}(\tau)\tilde{c}^\dagger_{m\uparrow}(0)
e^{-g(a_m^\dagger(0)-a_m(0))(1-\gamma\tilde{n}_{m\downarrow}(0))}>
\eeqn
We expand the exponentials in Eq. (30) using the operator relation
\beq
e^{g(a^\dagger-a)(1-\gamma \tilde{n}_\downarrow)}=
e^{g(a^\dagger-a)}+
\tilde{n}_\downarrow (e^{g(a^\dagger-a)(1-\gamma )}-e^{g(a^\dagger-a)})
\eeq
and decouple averages over bosons and fermions, following
Alexandrov and Ranninger. This leads to
\beqn
G(m,\tau)&=&\sigma_0(m,\tau)<-T\tilde{c}_{0\uparrow}
(\tau)\tilde{c}^\dagger_{m\uparrow}(0)>
\nonumber \\&+&[(\sigma_1(m,\tau)-\sigma_0(m,\tau)]
[<-T\tilde{n}_{0\downarrow}(\tau) \tilde{c}_{0\uparrow}
(\tau)\tilde{c}^\dagger_{m\uparrow}(0)>
+<-T \tilde{c}_{0\uparrow}(\tau)\tilde{c}^\dagger_{m\uparrow}(0)
\tilde{n}_{m\downarrow}(0)>]
\nonumber \\ &+&
[\sigma_2(m,\tau)-2\sigma_1(m,\tau)+\sigma_0(m,\tau)]
<-T\tilde{n}_{0\downarrow}(\tau) \tilde{c}_{0\uparrow}
(\tau)\tilde{c}^\dagger_{m\uparrow}(0)
\tilde{n}_{m\downarrow}(0)>
\eeqn
The boson Green's functions are defined as
\beq
\sigma_{i+j}(m,\tau)=
<e^{g(1-\gamma)^i(a_0^\dagger(\tau)-a_0(\tau))}
e^{-g(1-\gamma)^j(a_m^\dagger(0)-a_m(0))}>
\eeq
with $i,j=0,1$. At low temperatures they are given by\cite{mahan}
\bmath
\beq
\sigma_\alpha (m,\tau)=S^{2-\alpha}T^\alpha [1-\delta_{m,0}+
\delta_{m,0}e^{-g^2(1-\gamma)^\alpha D(\tau)}]
\eeq
\beq
D(\tau)=-[e^{-\omega_0|\tau|}+2\frac{cosh\omega_0\tau}{e^{\beta \omega_0}-1}]
\eeq
\emath
and in frequency space by
\beq
\sigma_\alpha(m,i\omega_n)=S^{2-\alpha}T^\alpha
[\delta_{n,0}\beta +\delta_{m,0}\sum_{l=1}^\infty 
\frac{2l\omega_0 g^{2l}(1-\gamma)^{\alpha l}}{l!(\omega_n^2+l^2\omega_0^2)}]
\eeq
We next decouple the fermion averages with the same mean field procedure 
used to calculate
the coherent part of the Green's function , and calculate the Fourier
transform
\beq
G(k,i\omega_n)=\int_0^\beta d\tau e^{i\omega_n \tau}\sum_me^{ikm}G_{0m}(\tau)
\eeq
The complete Green's function is
\beq
G(k,i\omega_n)=G_{coh}(k,i\omega_n)+G_{inc}(k,i\omega_n)
\eeq
For each term of the coherent Green's function, Eq. (18), there is
a corresponding term in the incoherent Green's function. All terms in
the coherent Green's function are of the form
\beq
G_{coh}^{\alpha,s}(k,i\omega_n)=
cS^{2-\alpha}T^\alpha \frac{a_k}{i\omega_n-sE_k}
\eeq
with $0\leq \alpha \leq 2$ and $s=+/-1$. The corresponding term
in the incoherent Green's function is
\beq
G_{inc}^{\alpha,s}(k,i\omega_n)=cS^{2-\alpha}T^\alpha 
\sum_{l=1}^\infty \frac{g^{2l}(1-\gamma)^{\alpha l}}{l!}
\frac{1}{N} \sum_{k'} a_{k'}[\frac{n_{k'}}{i\omega_n-s(E_{k'}-l \omega_0)}+
\frac{1-n_{k'}}{i\omega_n-s(E_{k'}+l \omega_0)}]
\eeq
The spectral function is obtained as usual from
\beq 
A(k,\omega)=-Im G(k,i\omega_n\rightarrow \omega+i\delta)
\eeq
and results from Eqs. (18), (37)-(39).  In particular, the 
lowest order normal
part of the incoherent spectral function is given by
\beqn
A_{inc}^n (k,\omega)&=&
S^2\times \sum_{l=1}^\infty \frac{g^{2l}}{l!}
[1+n[(1-\gamma)^le^{\gamma g^2}-1]]\nonumber \\
&\times&
\frac{1}{N}\sum_{k'}[u_{k'}^2[(1-n_{k'})\delta(\omega-l\omega_0-E_{k'})
+n_{k'}\delta(\omega+l\omega_0-E_{k'})]\nonumber \\
&+& v_{k'}^2[n_{k'}\delta(\omega-l\omega_0+E_{k'})
+(1-n_{k'})\delta(\omega+l\omega_0+E_{k'})]]
\eeqn
and the lowest order anomalous contribution by
\beqn
A_{inc}^a (k,\omega)&=&
S^2\times 2f_0 \times \sum_{l=1}^\infty \frac{g^{2l}}{l!}
[1+n[(1-\gamma)^le^{\gamma g^2}-1]]\nonumber \\
&\times&
\frac{1}{N}\sum_{k'} (-u_{k'}v_{k'})\times
[(1-n_{k'})\delta(\omega-l\omega_0-E_{k'})
+n_{k'}\delta(\omega+l\omega_0-E_{k'})\nonumber \\
&-& n_{k'}\delta(\omega-l\omega_0+E_{k'})
-(1-n_{k'})\delta(\omega+l\omega_0+E_{k'})]  .
\eeqn

\section{Results for the spectral function}
The spectral function for the models considered here is of the
form
\bmath
\beq
A(k,\omega)=A_{coh}(k,\omega)+A_{inc}(k,\omega)
\eeq
\beq
A_{coh}(k,\omega)=Z_h\delta(\omega-E_k)+Z_e\delta(\omega+E_k)
\eeq
\beq
A_{inc}(k,\omega)=-ImG_{inc}(k,\omega+i\delta)
\eeq
\emath
where the quasiparticle weights $Z_h$ and $Z_e$, and the incoherent
Green's function $G_{inc}$, were discussed in the previous two sections. 
As seen in Sect. IV, the quasiparticle weight $Z_e$ acquires a
$positive$ contribution from the onset of superconductivity. In a 
spectroscopic experiment usually only one side of the spectral function
is sampled, as the other side is suppressed by the Fermi function. The 
quantity that will display the enhanced coherence due to undressing
exhibited by $Z_e$ is
\beq
I_0(k,\omega)=A(k,\omega)f(\omega)
\eeq
with $f$ the Fermi function. In an experiment there will typically
be broadening from experimental resolution, which results in
\beq
I(k,\omega)=\int d\omega ' F(\omega - \omega ') I_0(k,\omega ')
\eeq
being measured, with $F(\omega)$ a Gaussian with width $\sigma_\omega$.
There could also be other sources of broadening of the $\delta$-functions
in the expressions Eq. (43) from lifetime effects. Just as the spectral
function, the measured spectrum Eq. (45) will have coherent and incoherent
contributions
\beq
I(k,\omega)=I_{coh}(k,\omega)+I_{inc}(k,\omega)
\eeq
arising from the coherent and incoherent parts of the spectral
function respectively.

Figure 13 shows results for the coherent spectra at the Fermi
energy for an underdoped ($n=0.02$, labeled $ud$),
optimally doped ($n=0.045$, labeled $op$) and overdoped
($n=0.1$, labeled $od$) case for the parameter values used in Sect. V.
For the underdoped case with the chemical potential below the bottom
of the band the value of $\epsilon_k$ at the bottom of the band was
used. The dashed lines show the spectra in the normal state at $T_c$, and
the full lines in the superconducting state at $T=0.1 T_c$. For each doping,
as superconductivity onsets the peaks shift to the left
due to the opening of the superconducting gap. Furthermore, the peaks
$grow$ in magnitude due to the behavior of $Z_e$ discussed in Sect. IV.
As function of doping the peaks grow in magnitude both in the normal
and superconducting state due to the enhanced coherence with increased
number of carriers. The superconducting peak in the $od$ case is shifted
to the right with respect to the $op$ case because the superconducting
gap is smaller in that region (Fig. 1(b)).

When we include the incoherent part of the spectra the smaller normal
state peak can become almost invisible. The results will depend
of course on the specific parameters chosen to describe the incoherent
background, and we are not suggesting that we are in a position to determine
them from first principles. In Figure 14 we show results for a particular 
set of parameters for the generalized Holstein model. In addition to the
parameters already discussed in Sect. IV, including the value of $\Upsilon$,
the new parameters needed are $S^2$, $\omega_0$ and a broadening factor, 
given in the figure caption. Note that in the underdoped case (a) the peak
in the normal state has become almost invisible, while a sharp peak and a
dip are seen in the superconducting spectrum. The dip arises because the
background term arising from the second term in Eq. (39) for
$a_{k'}=v_{k'}^2$, $s=-1$, is pushed to more negative energies as the
superconducting gap opens. As the doping increases the normal state
peak becomes more visible, and the overdoped case shows more conventional
behavior. Note that the scale in the figures changes with doping and the
magnitude of the peaks increases with doping.

Figure 15 shows the temperature dependence of the spectra for the overdoped 
case. The normal state peak is pushed back continuously as the 
superconducting gap opens up. In addition, for our case (a) the peak
grows in magnitude. To highlight the difference with conventional BCS
theory we show in Fig. 15(b) what is obtained with the same parameters
in the absence of the term $f_0$ in Eq. (18). The peak here first 
becomes lower and then increases again as the temperature is lowered,
but it is always lower or equal to the normal state peak. It is
easily seen from the BCS formula that this property is generally true 
also for other values of the momenta.

Similarly figure 16 shows the temperature dependence of the spectra
in the underdoped case. Here, rather than the peak moving
continuously, a new peak grows in the superconducting state. The presence of 
two peaks has not been seen experimentally in photoemission to our knowledge,
possibly because of  experimental resolution. For the BCS case (b) the
peak in the superconducting state is much smaller than in the normal
state, while for our case (a) the opposite is true. Results for the
temperature dependence in the optimally doped case show behavior
intermediate between the overdoped and underdoped cases shown.

\section{The cuprates}
We have seen in the previous sections that in systems where
superconductivity arises from undressing there is a signature
of the formation of the condensate in the single particle
spectral function. Specifically, it arises in Eq. (17) from the term
involving $f_0$, the on-site pair amplitude. Ding et al\cite{ding}
and Feng et al\cite{shen}, discussing
experimental results of angle resolved photoemission in cuprates have
recently emphasized precisely that feature of the observed spectra,
and correlated the growth of the peak in photoemission to quantities
related to the superconducting condensate such as the superfluid density
and the condensation energy. The spectra calculated within the present
theory in the previous section resemble in several aspects the 
experimental observations in photoemission along the 
$(\pi,0)$ direction.

Unfortunately, as the alert reader has undoubtedly noticed, the results
presented in the previous section with negative $\omega$ in a
hole representation correspond to $hole$ $destruction$, or $electron$
$creation$, that is, $inverse$ $photoemission$. It is for that case that
the experiment would sample $Z_e$, the quasiparticle weight for electron
creation. Instead, if we calculate spectra for direct photoemission
we would find that quasiparticle peaks are $suppressed$ by the onset
of superconductivity due to the behavior of $Z_h$ discussed in
Sect. IV.

The present theory does not allow for a switch in the role of the weights
$Z_e$ and $Z_h$: electron-hole asymmetry, of the sign assumed here,
is central to the theory. Does this then imply that the theory is
irrelevant for description of the cuprates?

We believe this is not the case. We propose that in fact, the
photoemission experiments along the $(\pi,0)$ direction close
to the $(\pi/a,0)$ point  sample
the part of the spectral function discussed in the previous section,
corresponding to hole destruction, or electron creation.

How can photoemission sample electron creation?
Recall that in the theory of hole superconductivity the relevant orbitals
are oxygen $p\pi$ orbitals in the planes\cite{holesc1}. There are however 
also the oxygen $p \sigma$ orbitals, strongly hybridized with the
Cu $d_{x^2-y^2}$ orbitals. Suppose that in a photoemission
experiment along the $(\pi,0)$ direction the largest matrix element
couples to destruction of a $d_{x^2-y^2}$ electron. This would not
directly couple to the band responsible for superconductivity; however,
that process could induce the destruction of an oxygen hole in the
$p \pi$ orbitals. The proposed situation is schematically depicted in Fig. 17,
in an electron representation. Before the photon comes in there is one
electron in each $Cu^{++}$ atom neighbor to a given $O$ atom, and two
holes in the $p\pi$ orbital on that $O$ atom. We assume the
energy level structure shown in Fig. 17: an electron from $Cu^{++}$
cannot 'fall' onto the $O$ $p\pi$ orbital because it is
Coulomb-repelled by the electron in the other $Cu$ atom. When a
photon comes in and knocks out one of the electrons in a $Cu$, the
other electron can 'fall' onto the $O$ $p\pi$ orbital, thus 
destroying an $O$ hole and sampling the quasiparticle weight $Z_e$.

It is clear that this qualitative explanation needs further
elaboration and experimental confirmation to be convincing.
Nevertheless we also point out that it suggests an explanation
for why the sharp peaks seen in photoemission along the
$(\pi,0)$ direction are not seen along the $(\pi,\pi)$ direction\cite{valla}:
since Cu $d_{x^2-y^2}$ orbitals point along the principal axis in the planar
square lattice, the coupling to the photon along the $(\pi,\pi)$ direction is
likely to be much smaller. For that direction the larger coupling may
be to the $O$ $p\pi$ band itself, in which case inverse
rather than direct photoemission would show the enhanced coherence.
It is possible that some indication of this effect may have already
been seen in tunneling experiments\cite{davis}.

\section{Conclusions}
In this paper we have continued exploring the consequences of the physical
principle proposed in I: that, in at least some electronic materials in nature,
the dressing of quasiparticle carriers is a function of the local
carrier concentration, and becomes smaller as the local carrier
concentration increases. This physical principle leads to superconductivity
occuring in these systems because of lowering of the carriers kinetic
energy upon pairing. The superconducting transition, and many of the
features of the superconducting state, have already been discussed earlier
within the theory of hole superconductivity\cite{holesc1,holesc2,strong}.

In this paper we explored the consequences of this principle for the single
particle Green's function in the superconducting state. The central result
of this paper, Eq. (18), demonstrates that formation of the superfluid
condensate will influence the behavior of the single particle
spectral function. Eq. (18) is thus the generalization of the
BCS spectral function for systems where superconductivity is driven
by undressing. Surprisingly, the results show that the enhanced coherence
in the superconducting state is displayed in the quasiparticle weight
for electron creation but not for electron destruction. We calculated
the behavior of the quasiparticle weights as function of temperature,
doping dependence and momentum, and highlighted their differences
with conventional BCS theory.

Furthermore, we discussed the calculation of the full spectral function
including the incoherent contribution for one particular model where
superconductivity occurs through undressing, a generalized Holstein 
model. Our calculation was performed within the Lang-Firsov 
approximation\cite{alex},
and it should be interesting to see whether the qualitative results survive
a more exact treatment\cite{marsiglio,white}.

Results for the full spectral function showed several features that resemble
experimental observations in photoemission experiments in high $T_c$ 
cuprates\cite{ding,shen}, in particular
enhanced coherence, as displayed by the quasiparticle peak in the spectra,
when the system enters the superconducting state and as the carrier
concentration increases both in the normal and in the superconducting state.

This study was strongly motivated by the beautiful experimental
results and insightful analysis of the photoemission experiments\cite{ding,shen}. 
Thus it is
perhaps disappointing that at the end of the day our calculation  predicts
these effects, in the simplest one-band model, arising in $inverse$ 
rather than in direct photoemission. Thus some readers may conclude that
our calculation is not more than an academic exercise. However, as
discussed in Sect. VII, we believe there is a plausible scenario by which
the spectral weight for electron creation would be sampled in the
photoemission experiments in the cuprates.

While the theory discussed here predicts s-wave rather than d-wave
superconductivity we believe it is remarkable how many of the features
that appear to be part of the phenomenology of
high $T_c$ cuprates it exhibits, as a
consequence of the  $single$ $assumption$ of a large value of the undressing
parameter $\Upsilon $:
(1) incoherence in the normal state at low hole concentration; (2) increased
coherence with doping in the normal state; (3) transition to superconductivity
for low doping, dissappearing for high doping, and bell-shaped $T_c$ versus hole
concentration; (4) increased coherence as the system goes superconducting;
(5) superconducting transition driven by kinetic energy lowering, optical
sum rule violation; (6) non-decrease of the quasiparticle gap at low
hole density when $T_c$ is going to zero. This latter feature arises in
our model from the fact that as the hole concentration decreases and the
band becomes narrower the chemical potential falls below the bottom
of the band\cite{strong}; we believe that many of the unusual properties of 
underdoped cuprates follow from this simple fact, and in particular that the
observed pseudogap is simply the energy difference between the
bottom of the band and the chemical potential.\cite{normal,coh}

If the theory of hole undressing discussed here describes the cuprates it 
is likely that it
is more generally applicable, because it is based on very general principles.
In this regard we note that one of the paradoxes of the conventional
explanation of superconductivity is that it is thought to
originate in an electron-boson (the electron phonon) coupling
that $opposes$ $conductivity$, i.e. gives rise to resistivity, in the normal state.
In a sense the present theory eliminates this paradox. Coupling to a boson
is certainly necessary, and that coupling gives rise to enhanced resistivity
in the normal state due to enhanced effective mass, but superconductivity
arises from a process whereby the coupling to that boson is $reduced$
as carriers pair and the system becomes superconducting. The old paradox is 
however replaced by a 
new one, that in order for heavily dressed 'confined' carriers to 
become less dressed, or 'freer', it is necessary for them to $bind$ in
Cooper pairs.

We also note that the principle on which the present theory is based, that
an increase in the local hole occupation causes undressing, is likely to
be more general than as expressed by Eq. (2): rather than just by
the same site occupation, undressing may also be enhanced by hole
occupation of neighboring sites, and also by neighboring bond occupation.
The possible implications of this for superconductivity and other
instabilities of metals will be discussed in future work.

If indeed the essential physics of high $T_c$ is hole undressing,
what makes a material a high $T_c$ superconductor? Presumably,
the fact that quasiparticles are heavily dressed in the normal
state together with the fact that the undressing process that
occurs when the local carrier concentration increases is particularly
efficient. Both of those facts are  necessary
conditions for high $T_c$ superconductivity by giving rise
to a large $\Upsilon $ parameter. We will not discuss here what 
aspects of the chemistry
of the cuprates would favor this situation\cite{diat}.
However, conversely, we may conclude that the reason for a material
$not$ being a high $T_c$ superconductor would be a small value
of the parameter $\Upsilon $, either because
quasiparticles are $not$ heavily dressed in the normal state (e.g. the
case of Aluminum),
or, because the quasiparticle dressing in the normal state may not be 
strongly dependent on the local carrier concentration (e.g. the case
of 'heavy fermion' systems).

\acknowledgements
The author is grateful to F. Driscoll for the donation of a computer where
the calculations reported here were performed.

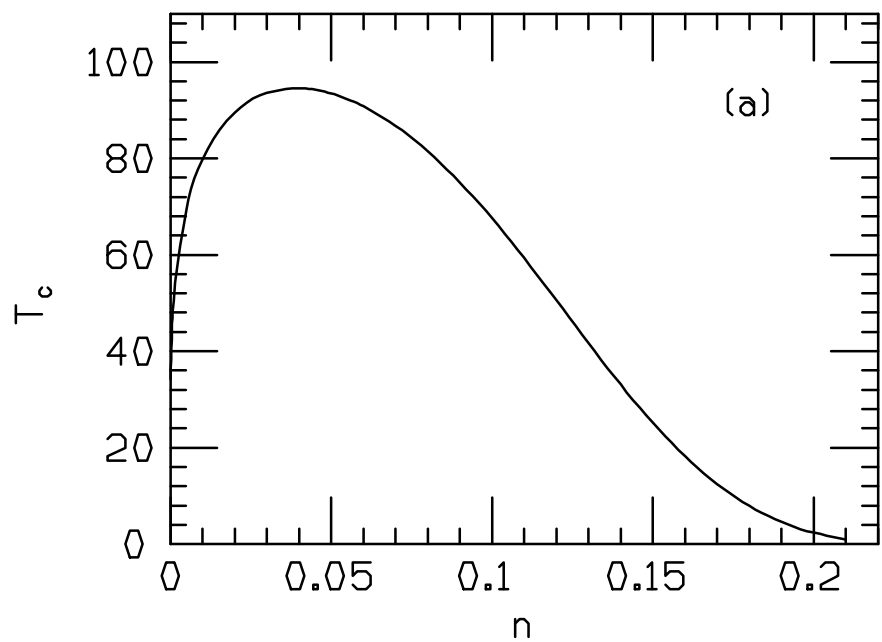
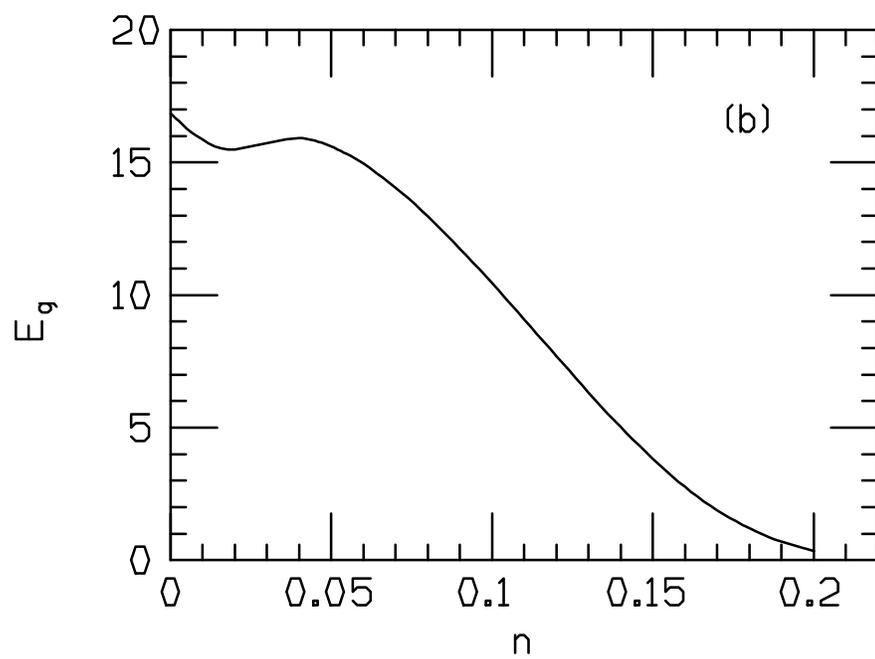
\begin{figure}
\caption {
 Superconducting transition temperature versus hole
doping $n$, number of holes per planar oxygen, for parameters given
by Eq. (25). (b) Minimum quasiparticle excitation energy at
low temperatures versus doping.}
\label{Fig. 1}
\end{figure}

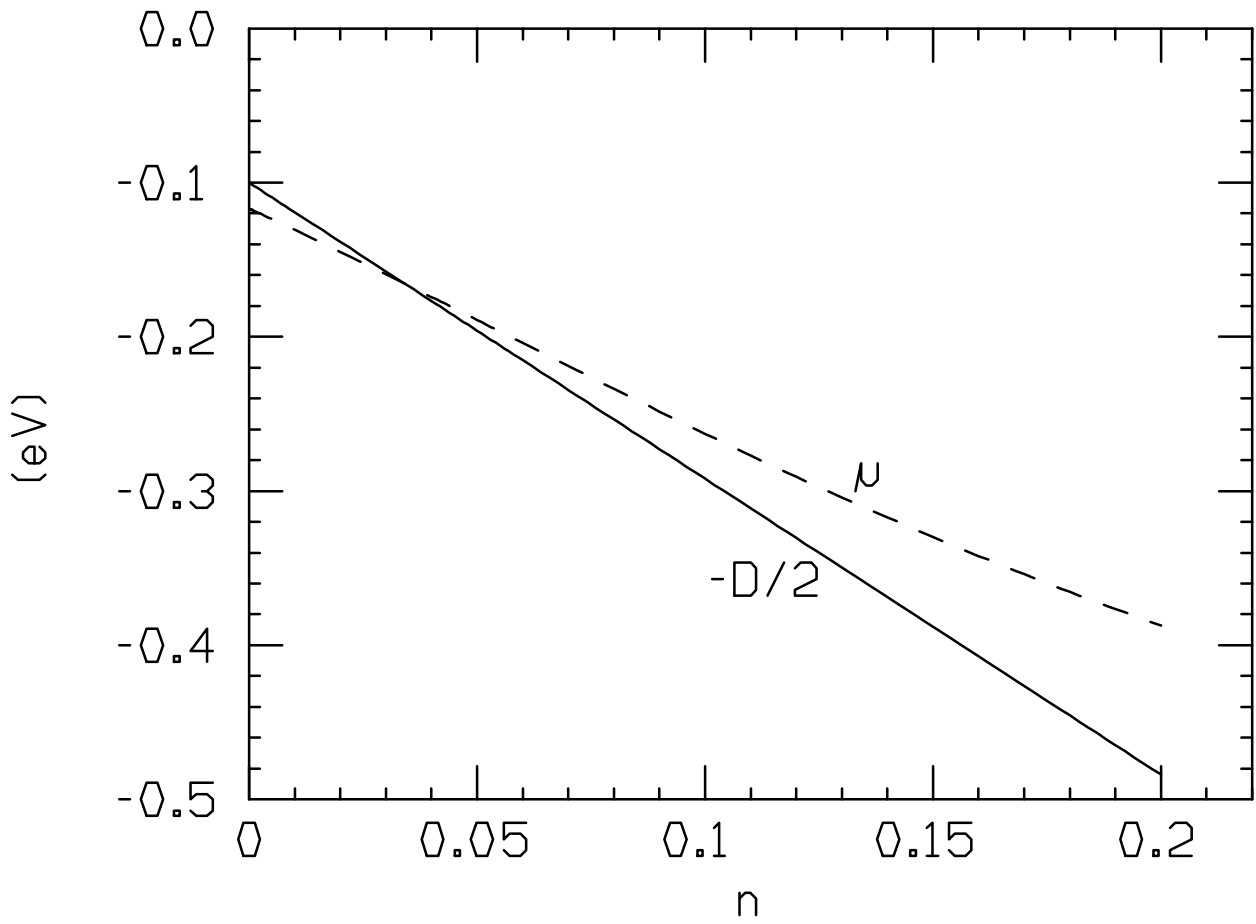
\begin{figure}
\caption {Chemical potential $\mu $ at low temperatures,
and band bottom $(-D/2)$ versus hole doping
for the parameters Eq. (25).
The chemical potential falls below the bottom of the band for hole
concentration $n \sim 0.038$.  For fixed hole
concentration $\mu $ increases as the  temperature is lowered
above $T_c$, particularly for low hole concentration, and stays approximately 
constant below $T_c$ for all hole concentrations.
}\label{Fig. 2}
\end{figure}

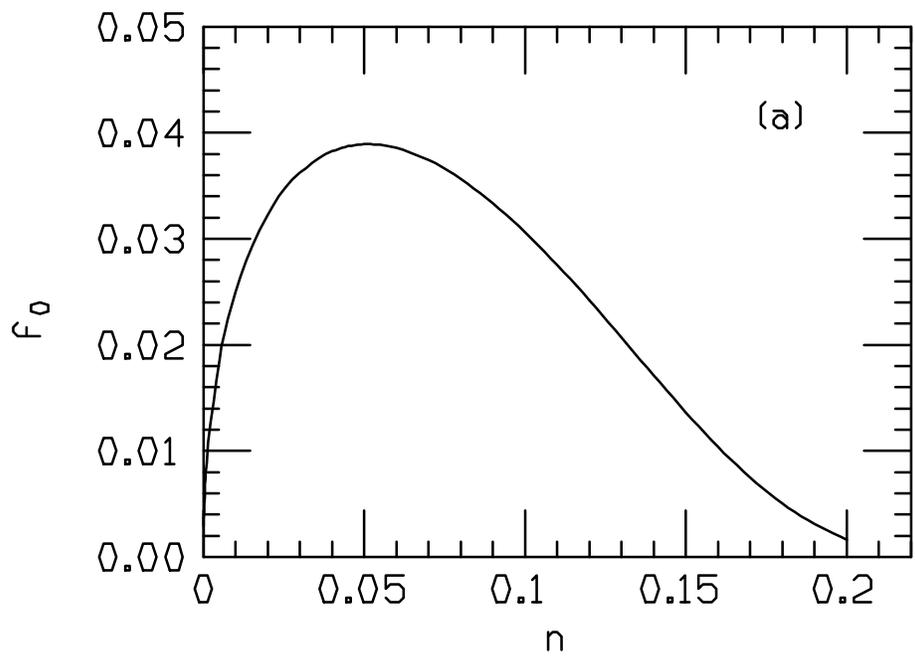
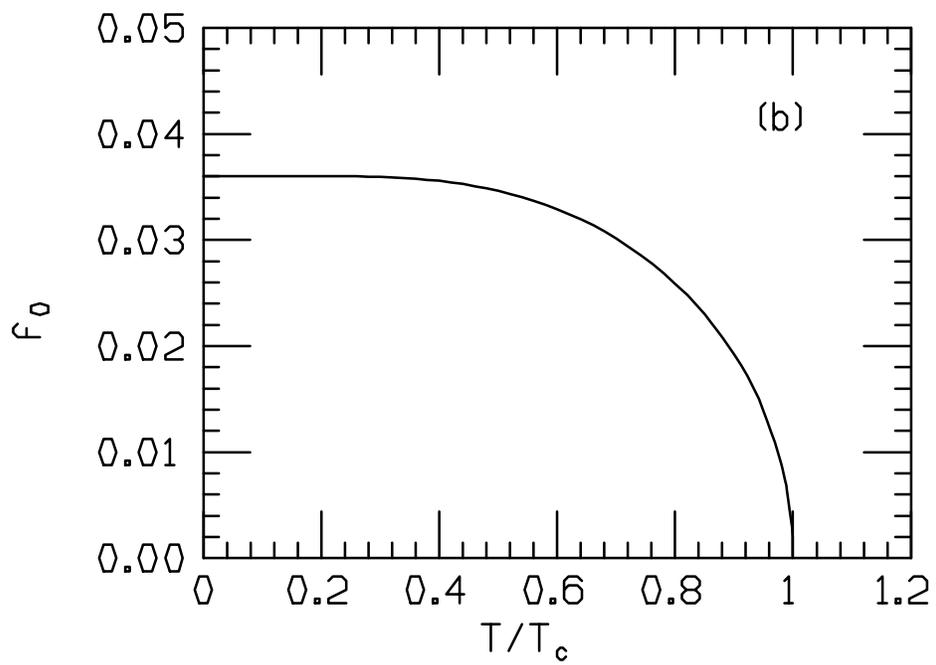
\begin{figure}
\caption {On-site pair amplitude $f_0$ (a) versus hole doping at low
temperatures, and (b) versus temperature at optimal doping.
}\label{Fig. 3}
\end{figure}

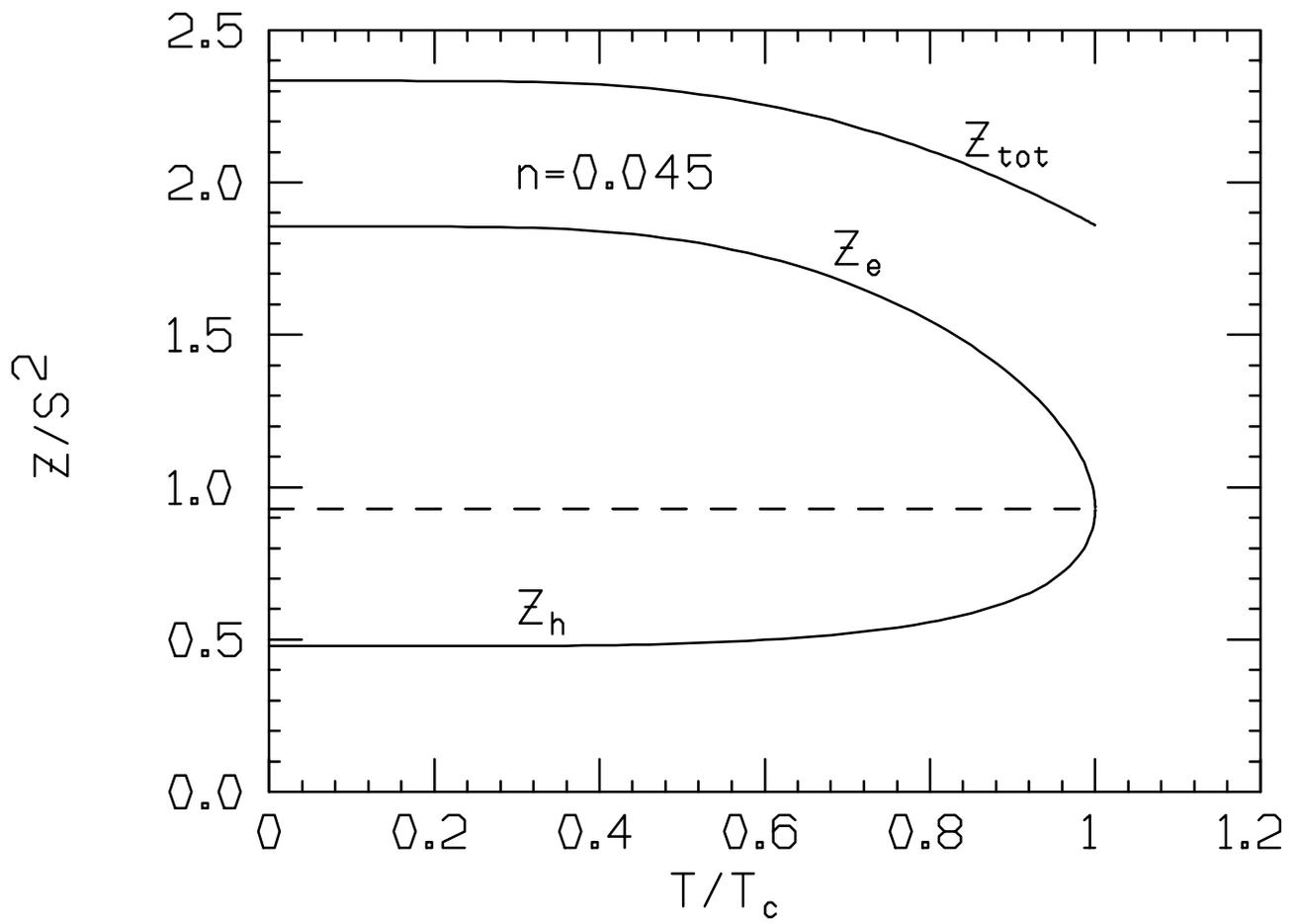
\begin{figure}
\caption {Quasiparticle weights at the Fermi energy ($\epsilon_k=\mu$)
versus temperature for the optimally doped case. $T_c=94K$.  $Z_{tot}$ is the
sum of $Z_h$ and $Z_e$. The corresponding BCS results are equal to
each other and independent of temperature (dashed line).
}\label{Fig. 4}
\end{figure}

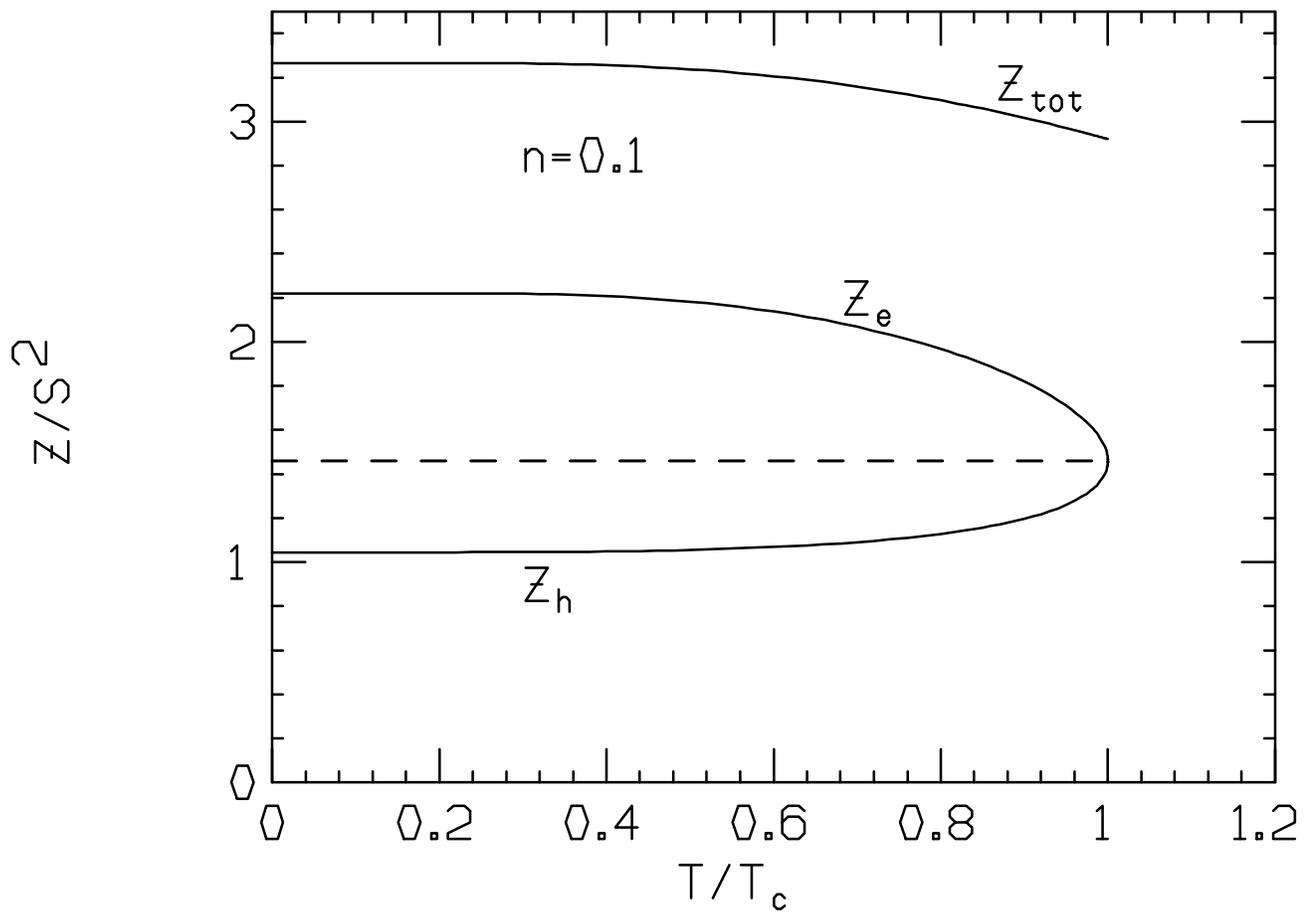
\begin{figure}
\caption {Same as Fig. 4 for an overdoped case, $n=0.1$, with $T_c=68K$.
}\label{Fig. 5}
\end{figure}

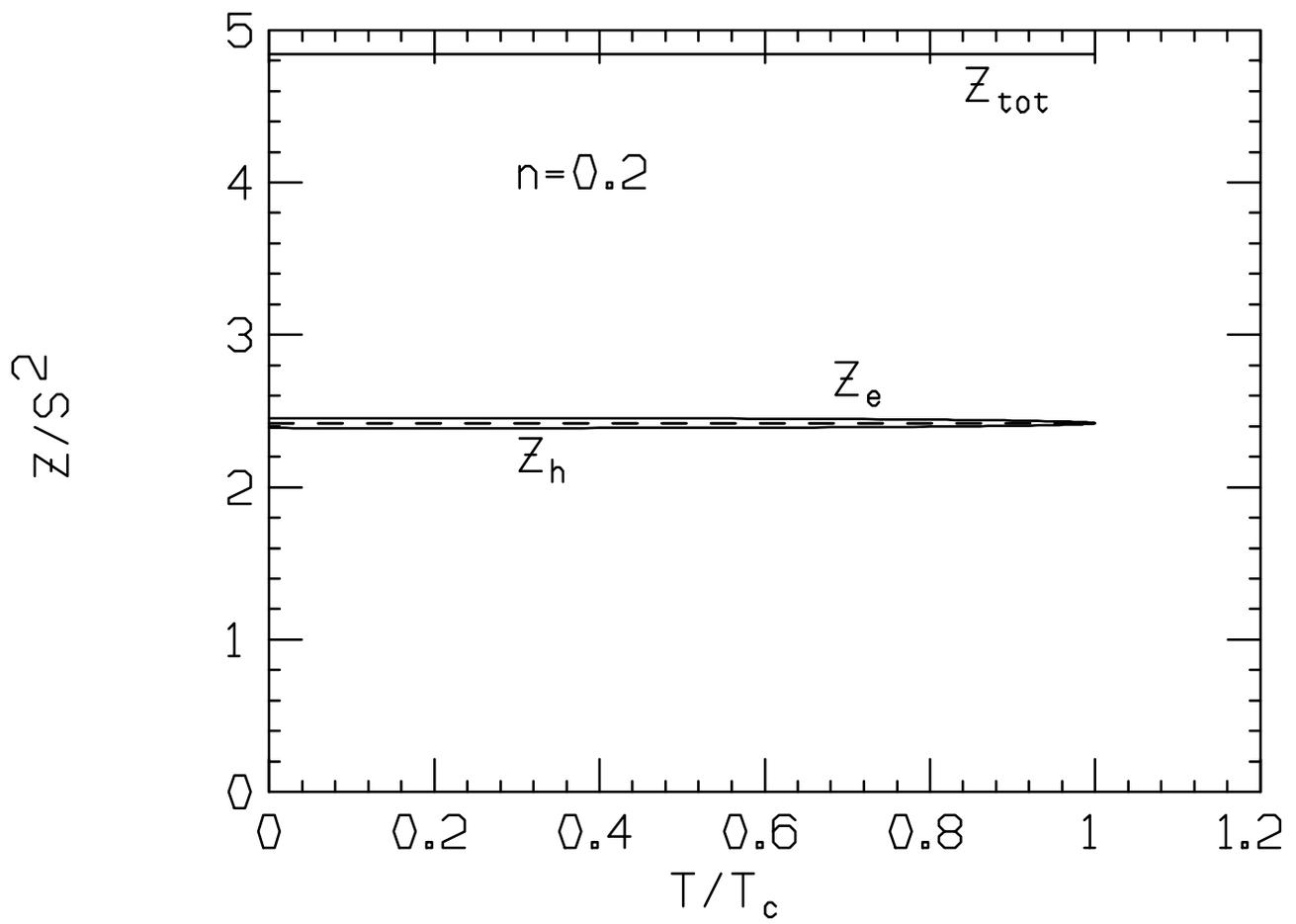
\begin{figure}
\caption {Same as Fig. 4 for a highly overdoped case, $n=0.2$,
with $T_c=2.4K$.
}\label{Fig. 6}
\end{figure}

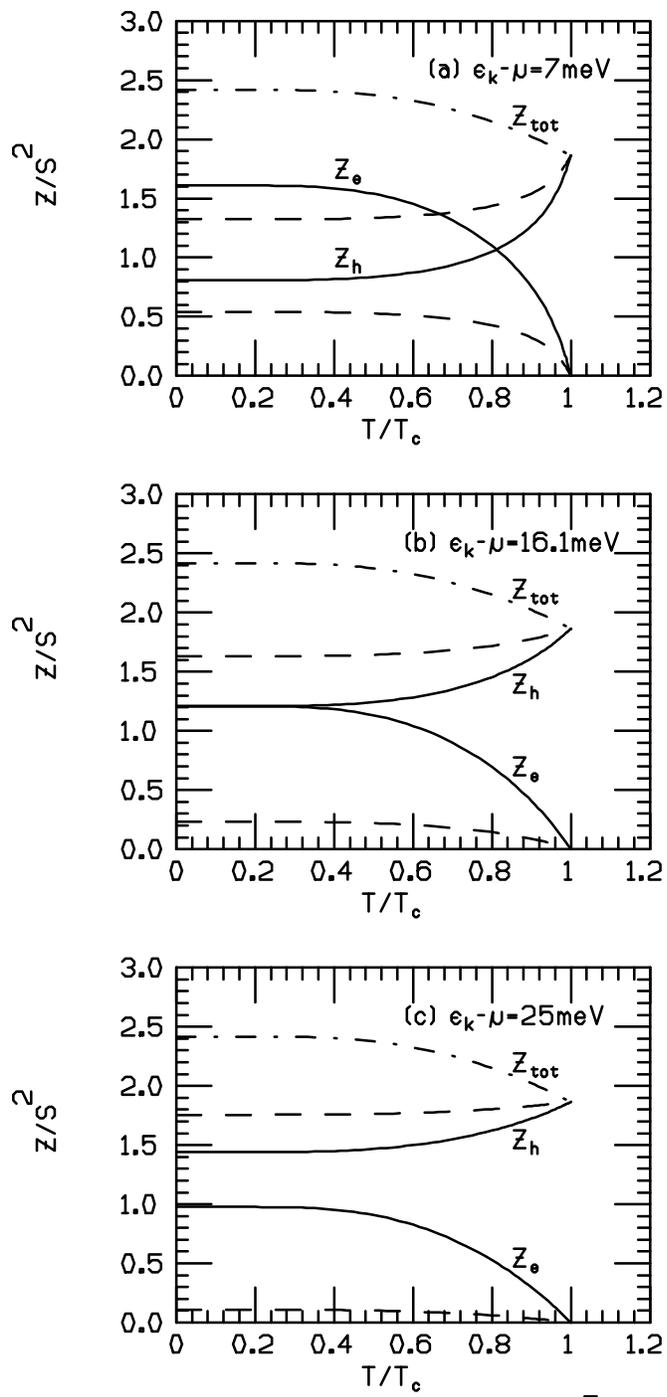
\begin{figure}
\caption {Spectral weights at optimal doping versus temperature
for momentum $inside$ the $electron$ Fermi surface. The 
dashed lines give the BCS values (Eq. (18) with 
$f_0=0$, $u_k^2\sim Z_h$, $v_k^2\sim Z_h$).
The upper dashed line corresponds to $u_k^2$, the lower one to $v_k^2$.
}\label{Fig. 7}
\end{figure}

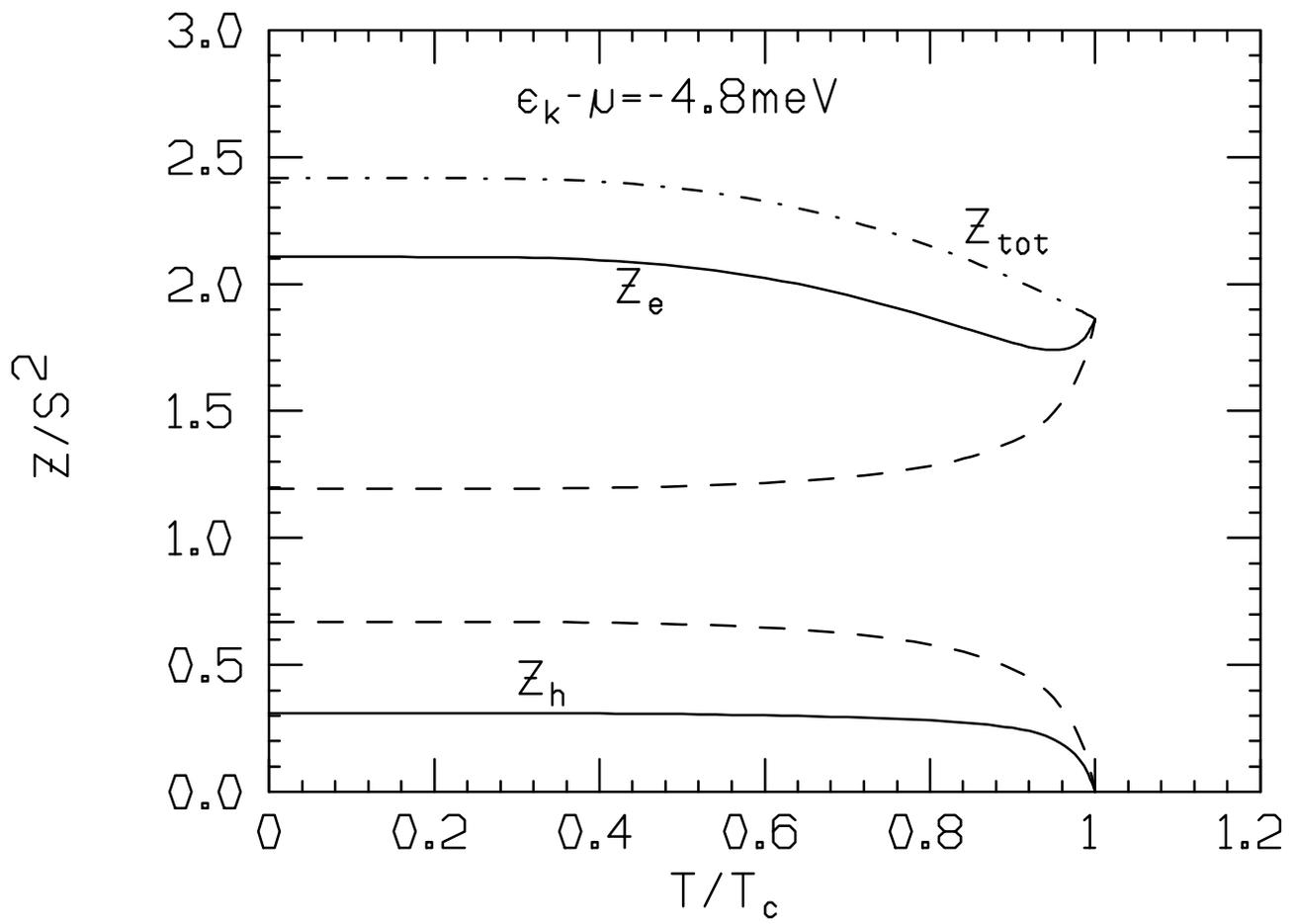
\begin{figure}
\caption {Spectral weights at optimal doping versus temperature
for momentum $outside$ the $electron$ Fermi surface. The 
dashed lines give the BCS values,
the upper dashed line corresponds to $v_k^2$, the lower one to $u_k^2$.
}\label{Fig. 8}
\end{figure}

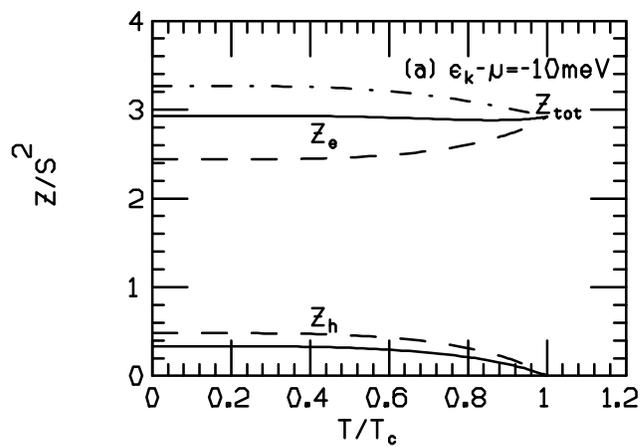
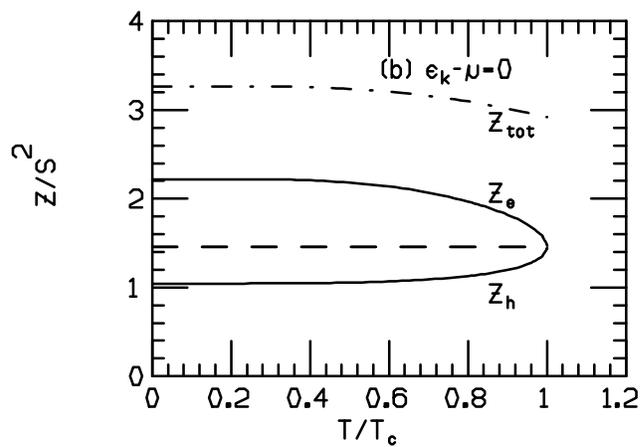
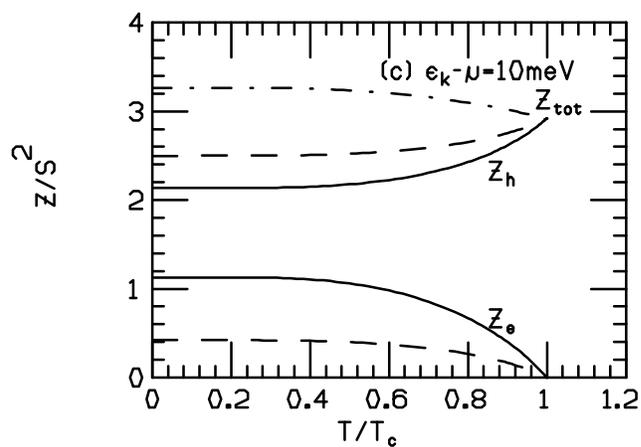
\begin{figure}
\caption {Spectral weights versus temperature for an overdoped
case, $n=0.1$, and momentum (a) outside, (b) at, and (c) inside
the electron Fermi surface.
}\label{Fig. 9}
\end{figure}

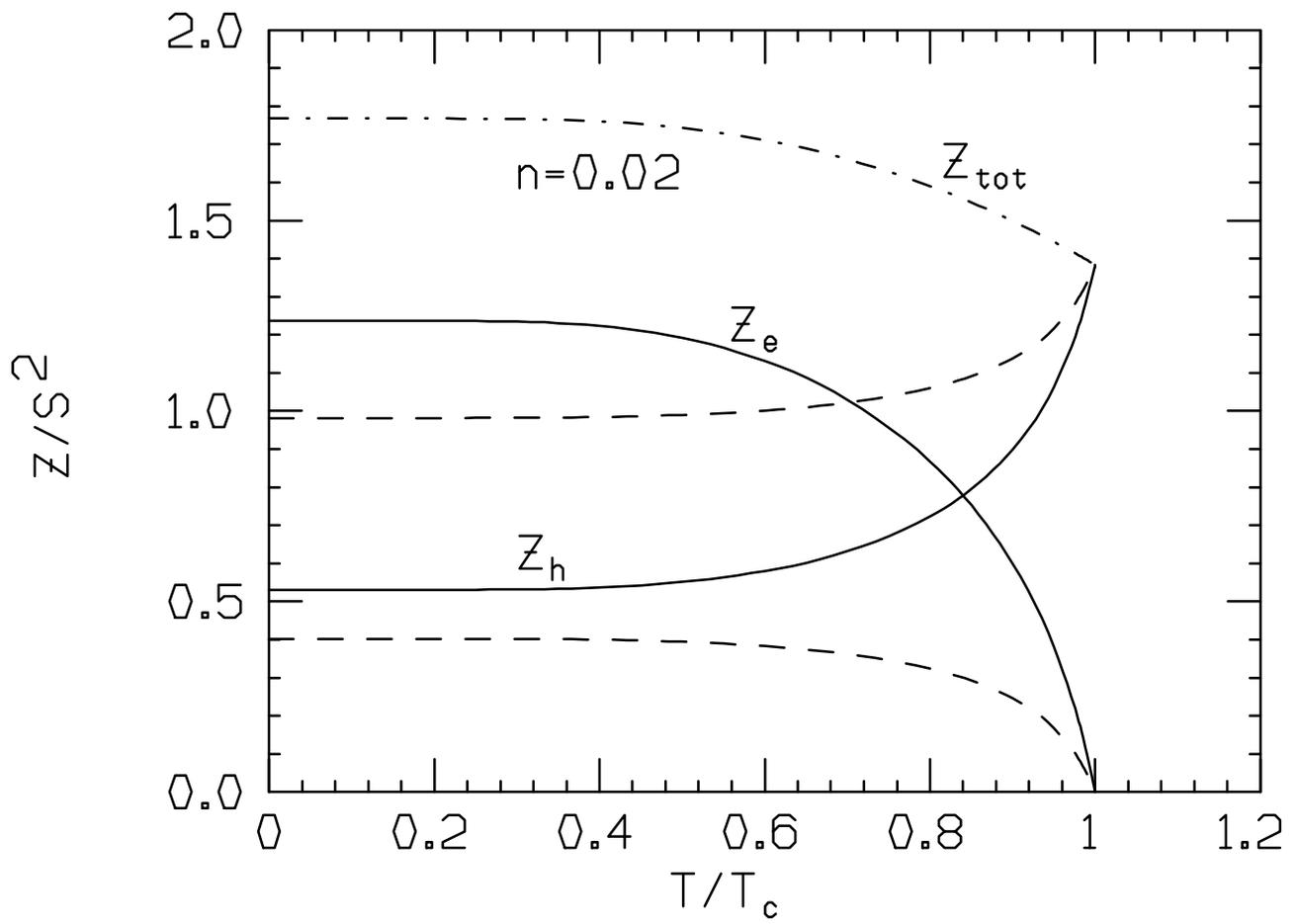
\begin{figure}
\caption {Spectral weights versus temperature for an underdoped
case, $n=0.1$, and momentum  inside
the electron Fermi surface. The value of $\epsilon_k-\mu=6.5 meV $ 
corresponds to $\epsilon_k$ at the bottom of the hole band.
}\label{Fig. 10}
\end{figure}

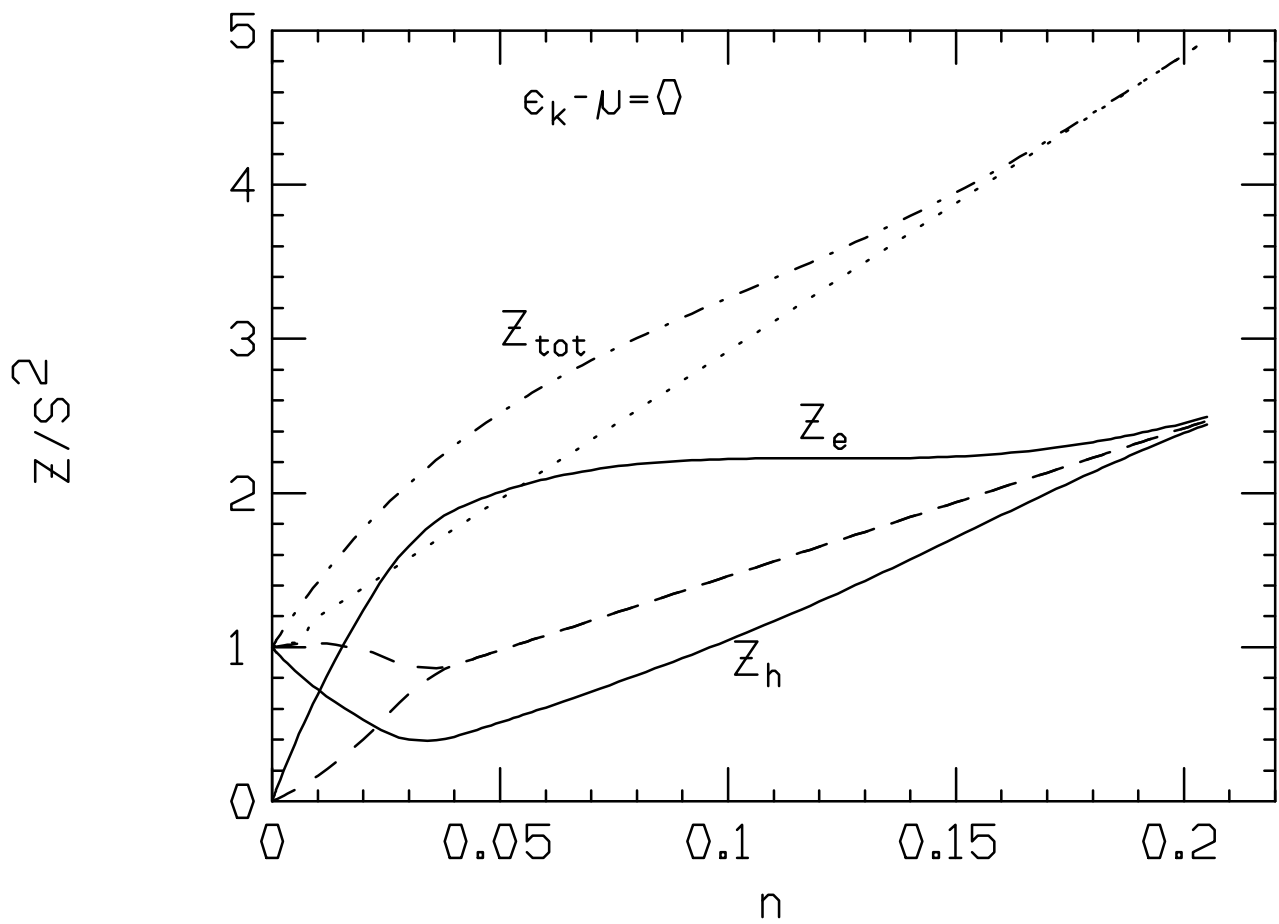
\begin{figure}
\caption {Quasiparticle weights versus doping at low temperatures.
The weights are computed at the chemical potential when it is inside
the band, and at the lower hole band edge in the underdoped regime when
the chemical potential is below the band edge. The lower dashed line
gives the BCS values $u_k^2$ and $v_k^2$, which are equal when
$\mu$ is inside the band and separate into two (upper corresponding to
$u_k^2$,lower to $v_k^2$) when $\mu$ is below the hole band edge.
The dot-dashed line $Z_{tot}$ gives the total weight $Z_h+Z_e$ and
the dotted line close to it the corresponding BCS total weight. 
Note that $Z_e$ rises approximately linearly with doping for
low hole doping and levels off in the overdoped regime. All results
approach the BCS values for high doping as $f_0$ approaches zero,
but remain different from the BCS values as long as $T_c$ is nonzero.
}\label{Fig. 11}
\end{figure}

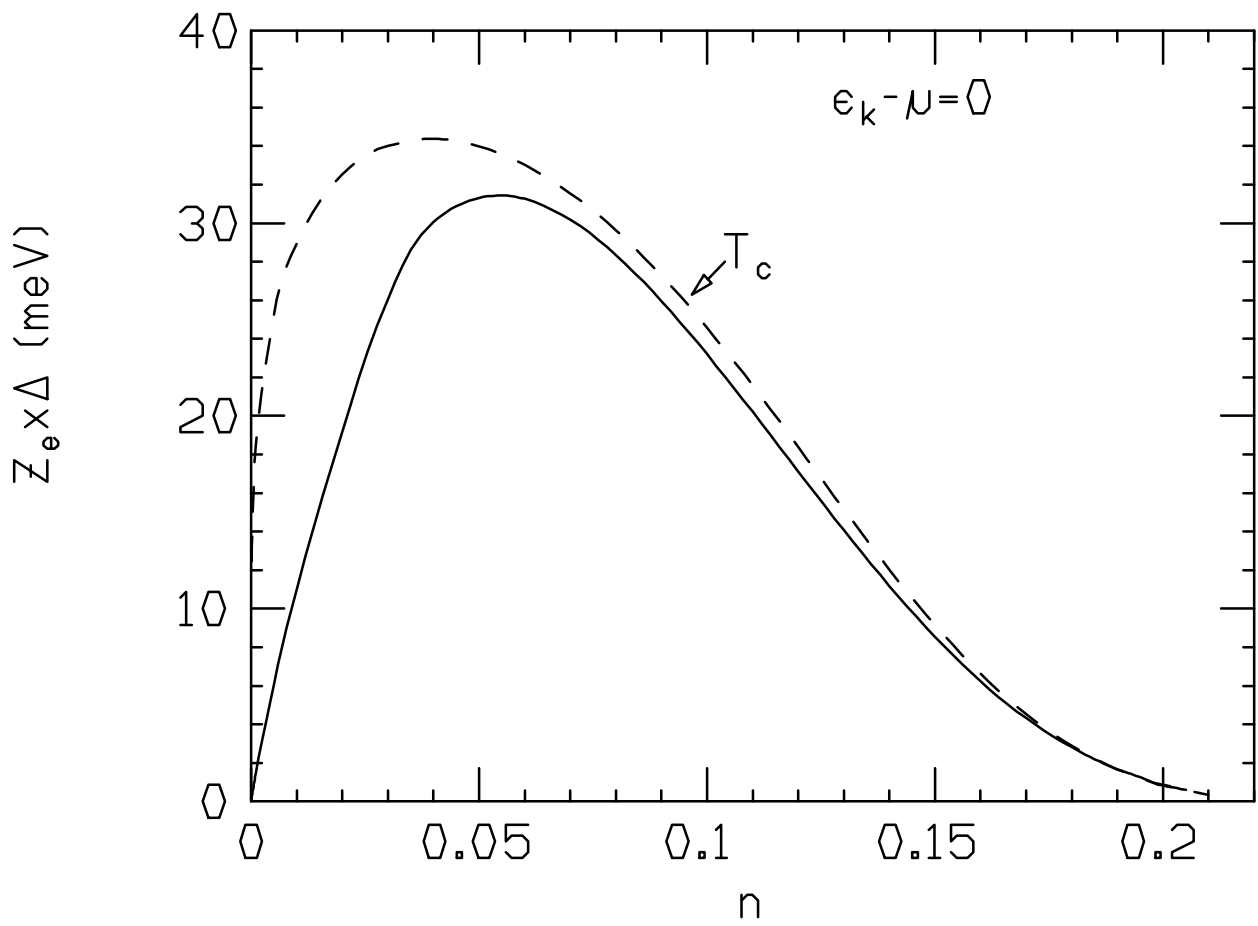
\begin{figure}
\caption {Product of electron quasiparticle weight $Z_e$ and minimum
quasiparticle excitation energy $E_g$ at low temperatures versus hole doping.
The dashed line gives $T_c$ versus doping. Note that $Z_e \times \Delta $
peaks at somewhat higher values of hole doping than $T_c$ does.
}\label{Fig. 12}
\end{figure}

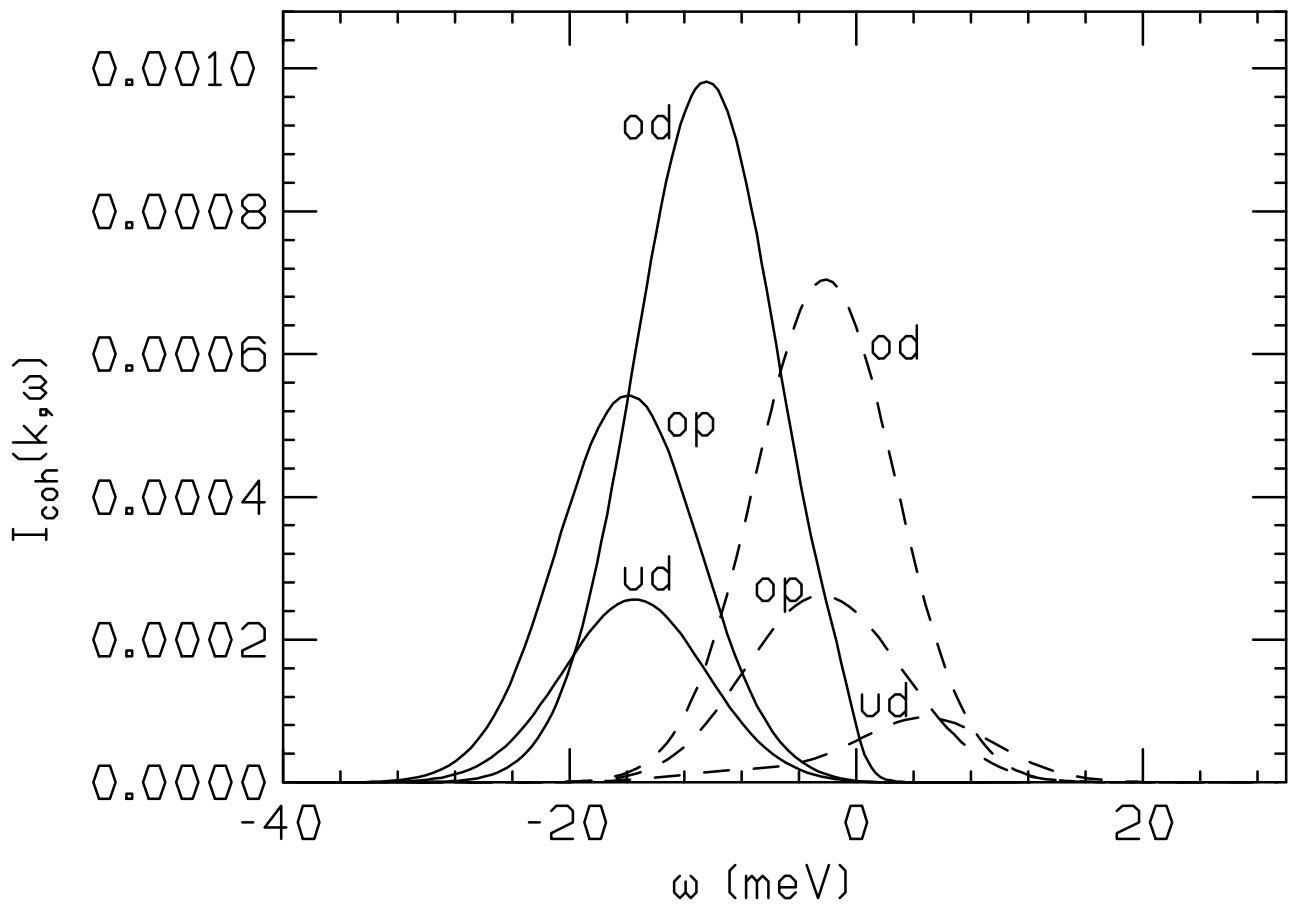
\begin{figure}
\caption {Coherent part of the spectral function multiplied by the
hole Fermi function $f(\omega)$ and broadened by a Gaussian function,
Eq. (43), with $\sigma_\omega =5 meV$. The dashed lines give the 
results at $T_c$, the full lines the results at low temperatures ($T=0.1 T_c$).
$od$, $op$ and $ud$ denote overdoped ($n=0.1$), optimally doped
($n=0.045$) and underdoped ($n=0.02$) regimes.
}\label{Fig. 13}
\end{figure}

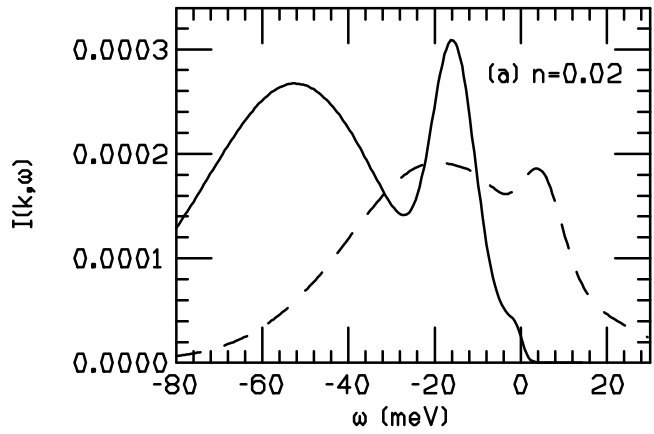
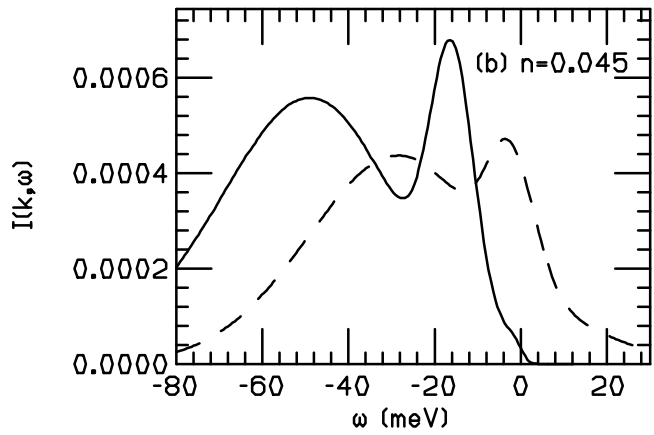
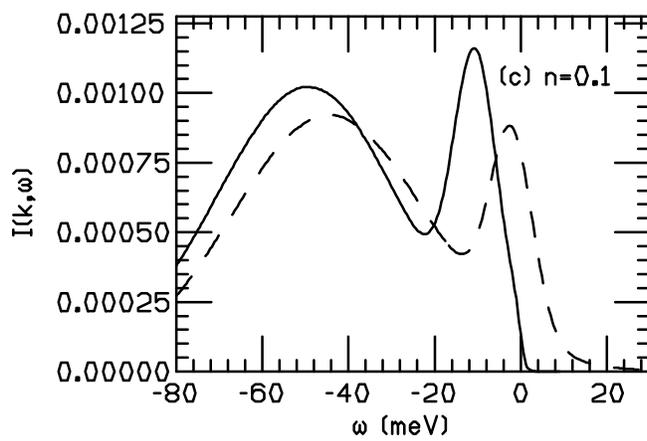
\begin{figure}
\caption {Results for the full spectra Eq. (43) at $T=T_c$ (dashed lines)
and at low temperatures, $T=0.1 T_c$ (solid lines). The incoherent part
of the spectrum was modeled with a generalized Holstein model
with $\omega_0=5meV$ and gaussian broadening for the $\delta$
functions $\sigma=15meV$. The band narrowing parameter is $S^2=1/500$,
and $\Upsilon=19.2$, corresponding to $g=2.49$ and $\gamma=0.45$ in Eq. (26).
The momentum is given by the normal state Fermi momentum for the optimally
doped and overdoped cases, and by the momentum corresponding to
the bottom of the hole band for the underdoped case.
}\label{Fig. 14}
\end{figure}

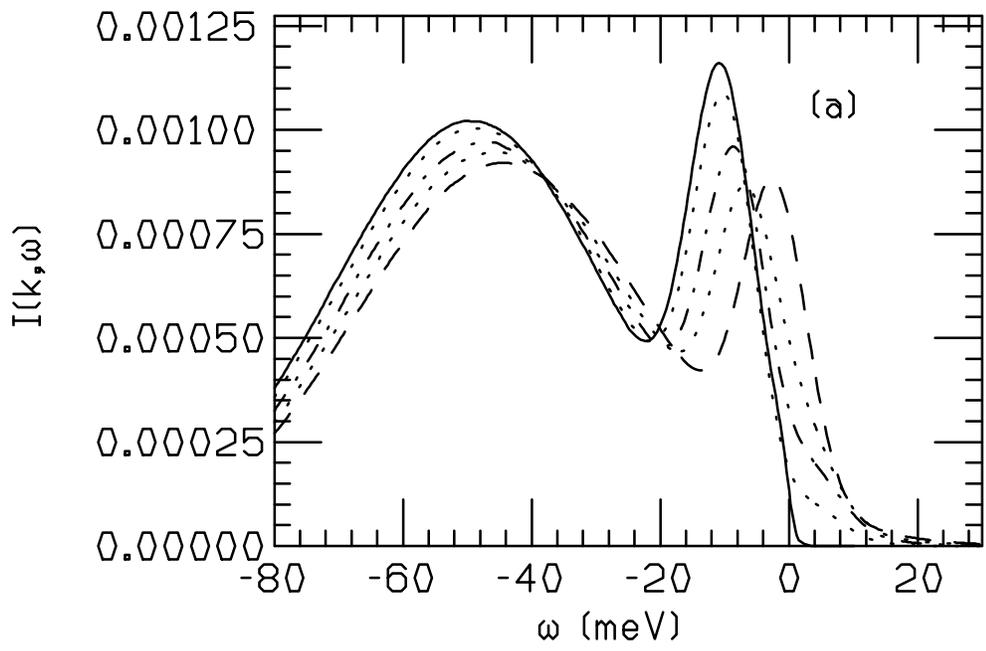
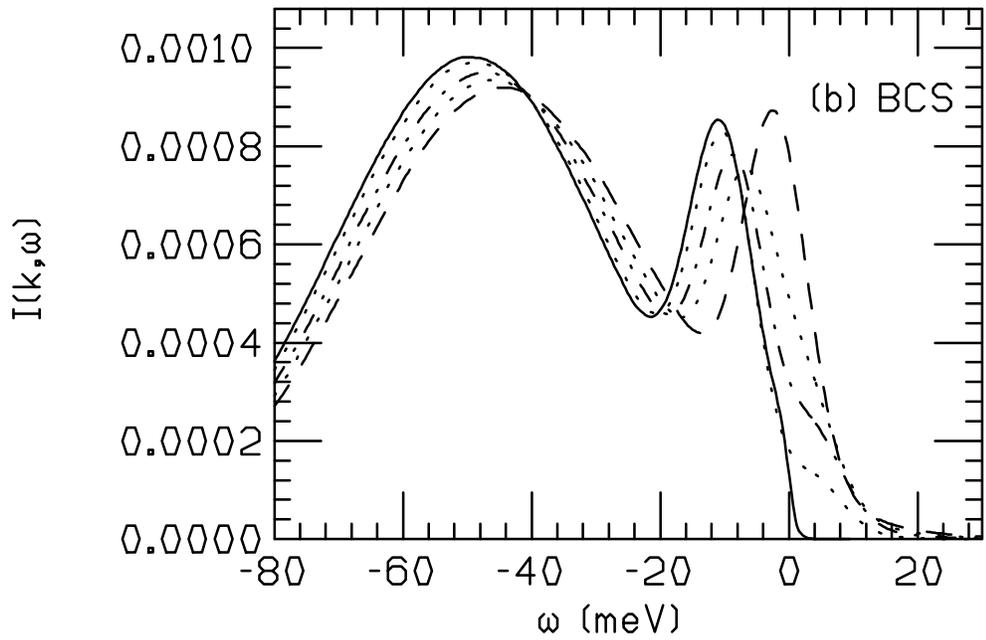
\begin{figure}
\caption {Spectra in the overdoped case $n=0.1$ at the chemical
potential for various temperatures. The dashed, dotted, dot-dashed,
dotted and full lines, with the peak moving towards the left,
correspond to $T/T_c=1, 0.9, 0.8, 0.6$ and $0.1$ respectively. 
In (b) the corresponding results for the BCS case, taking
$f_0=0$ in Eq. (18), are shown. Note that the peaks in the
superconducting state are always lower than that in the normal
state in the BCS case.
}\label{Fig. 15}
\end{figure}

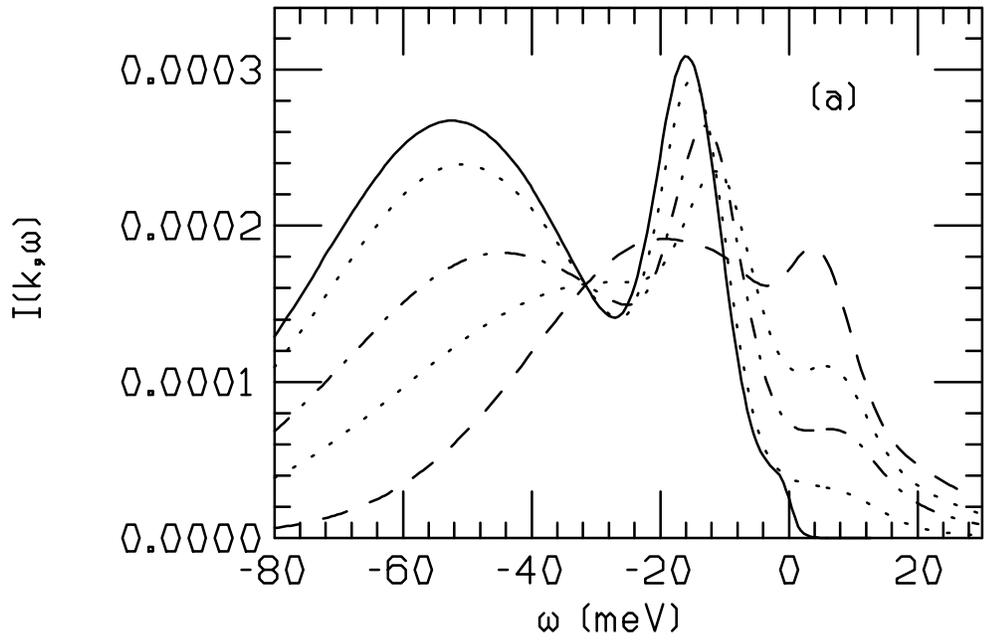
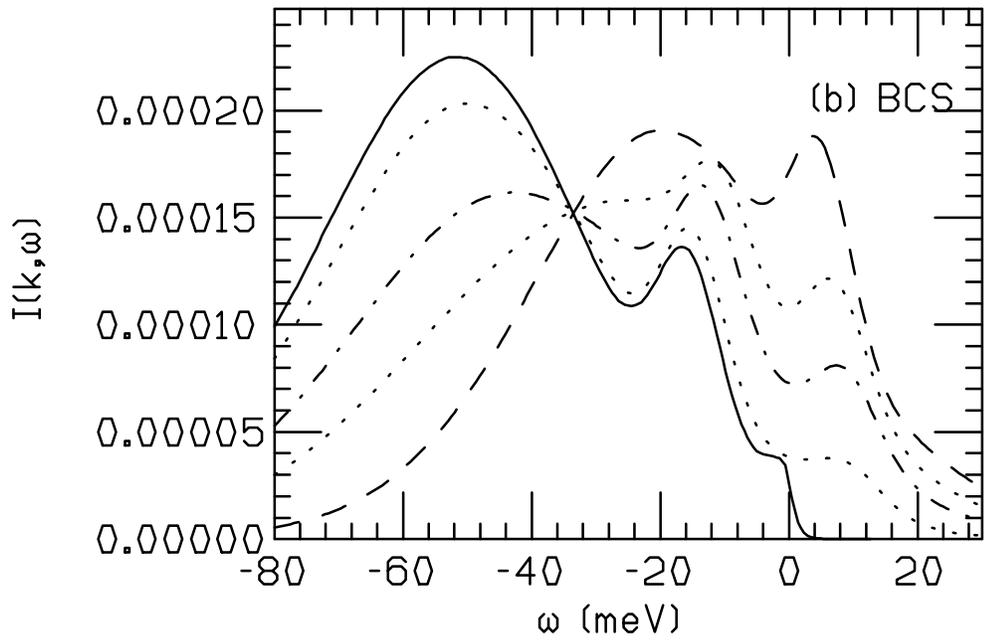
\begin{figure}
\caption {Same as Fig. (15) for an underdoped case, $n=0.02$, for
momentum corresponding to the bottom of the hole band. The same line
convention as in Fig. (15) is used.
}\label{Fig. 16}
\end{figure}

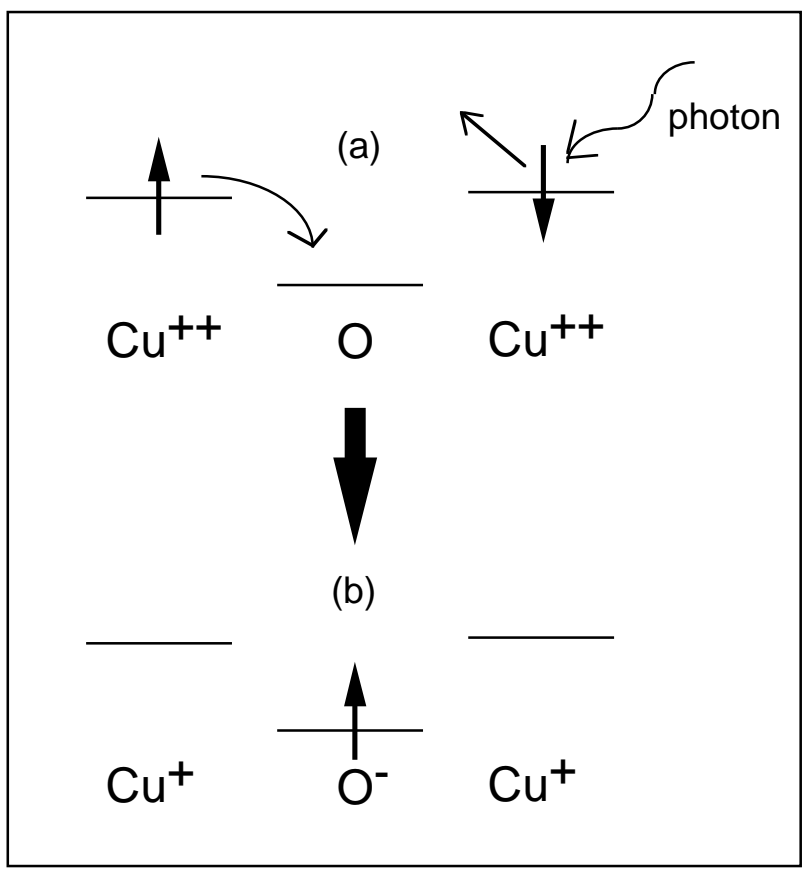
\begin{figure}
\caption {Schematic proposed explanation of how electron extraction in
a photoemission experiment can give rise to hole destruction
in the oxygen band responsible for superconductivity. The $Cu$
energy level arises from a hybridized $Cu$ $d_{x^2-y^2}$-$O$ $p\sigma$
orbital, the $O$ energy level corresponds to an $O$ $p\pi$ planar orbital.
In (a), an electron in the $Cu$ $d_{x^2-y^2}$ orbital is
knocked out by an incoming photon; the electron from a neighhboring
$d_{x^2-y^2}$ orbital then falls onto the $O$ $p\pi$ orbital (b),
destroying an $O$ hole.
}\label{Fig. 17}
\end{figure}

\end{document}